\documentclass[11pt,a4paper]{article}

\pdfoutput=1
\usepackage{jcappub}
\usepackage{graphicx}
\usepackage{amssymb}
\usepackage{amsmath}
\usepackage{mathtools}
\usepackage{bm}
\usepackage{latexsym}
\usepackage{epsfig}
\usepackage{textcomp}
\usepackage{color}
\usepackage{float}
\usepackage{blindtext}
\usepackage{enumitem}
\usepackage{xcolor}
\usepackage[section]{placeins}
\usepackage{caption}
\usepackage{float}
\usepackage{siunitx}
\usepackage{tikz} 
\usepackage{pgfplots}
\usepackage{subcaption}
\usepackage[labelformat=parens,labelsep=quad,skip=3pt]{caption}
\usepackage{graphicx}
\pgfplotsset{compat=newest} 
\usepgfplotslibrary{units} 

\usepackage{calligra}
\DeclareMathAlphabet{\mathcalligra}{T1}{calligra}{m}{n}
\DeclareFontShape{T1}{calligra}{m}{n}{<->s*[2.2]callig15}{}

\usepackage[mathscr]{euscript}

\makeatletter
\gdef\@fpheader{}
\makeatother

\def\l{\left}
\def\r{\right}

\def\beq{\begin{equation}}
\def\eeq{\end{equation}} 
\def\be{\begin{eqnarray}}
\def\ee{\end{eqnarray}}

\title{Primordial black holes dark matter from inflection point models of inflation and the effects of reheating}
\author{Nilanjandev Bhaumik and}
\author{Rajeev Kumar Jain}
\affiliation{Department of Physics, Indian Institute of Science, Bangalore 560012, India}
\emailAdd{nilanjandev@iisc.ac.in}
\emailAdd{rkjain@iisc.ac.in}

\abstract
{We study the generation of primordial black holes (PBH) in a single field inflection point model of inflation wherein the effective potential is expanded up to the sextic order and the inversion symmetry is imposed such that only even powers are retained in the potential. Such a potential allows the existence of an inflection point which leads to a dynamical phase of ultra slow roll evolution, thereby causing an enhancement of the primordial perturbation spectrum at smaller scales. Working with a quasi-inflection point in the potential, we find that PBHs can be produced in our scenario in the asteroid-mass window with a nearly monochromatic mass fraction which can account for the total dark matter in the universe. For different choices of quasi-inflection points and other parameters of our model, we can also generate PBHs in higher mass windows but the primordial spectrum of curvature perturbations becomes strongly tilted at the CMB scales. Moreover, we study the effects of a reheating epoch after the end of inflation on the PBHs mass fraction and find that an epoch of a matter dominated reheating can shift the mass fraction to a larger mass window as well as increase their fractional contribution to the total dark matter even for the case of a monochromatic mass fraction.}

\arxivnumber{}

\begin{document}
\maketitle


\section{Introduction}

The recent detection of astrophysical gravitational waves (GW) emitted from a system of binary black holes (BH) has reinitiated an immense interest in exploring the possibility that such BHs could also constitute a significant fraction of the cold dark matter (CDM) in our universe. A few such events have been reported in a short span of time by the LIGO-Virgo scientific collaboration \cite{Abbott:2016blz, Abbott:2016nmj, Abbott:2017vtc, Abbott:2017oio}. A generic and interesting conclusion that emerges from the analysis of the LIGO-Virgo data is that these BHs are very massive, with masses $\gtrsim {\cal O} (10\, M_{\odot})$ where $M_{\odot} \sim 2\times 10^{33}\, {\rm g}$ \cite{Abbott:2016htt, Belczynski:2016obo}. As a result, it was pointed out that these BHs could be primordial {\it i.e.} produced in the very early universe \cite{Hawking:1971ei, Carr:1974nx, Khlopov:1985jw, Ivanov:1994pa, Duechting:2004dk, Bean:2002kx}. When cosmological fluctuations with large overdensities exit the horizon during inflation and re-enter during the radiation  dominated epoch, they would collapse rather quickly and form these primordial black holes (PBH). Since the PBHs are produced before the big bang nucleosynthesis (BBN), they are considered non-baryonic, non-relativistic and effectively collisionless and thus could be a very promising  candidate for the CDM (or at least a fraction of it !) in the universe \cite{Bird:2016dcv, Clesse:2016vqa, Sasaki:2016jop, Carr:2016drx}. PBHs are therefore widely considered a very unique non-baryonic candidate for the CDM which remain free from the BBN constraints on the total baryonic density in the universe.

Although PBHs are an interesting class of candidates for the CDM in the universe, their mass spectrum is nevertheless strongly constrained by several different observations such as CMB anisotropies, capture of PBHs by neutron stars and gravitational femtolensing. For a monochromatic mass spectrum which is appropriate for a narrow mass distribution, these observations constrain different mass ranges for PBHs as we have briefly discussed below:

\begin{enumerate}
\item
$M \sim 10^{-18}  - 10^{-16} M_{\odot}$: 
This constraint (on the smallest mass range) simply comes from the non-detection of the evaporation signatures of PBHs in the extragalactic photon background. Due to Hawking radiation\footnote{In higher dimensional theories, the presence of extra spatial dimensions affects how gravity acts on small scales and could slow down the Hawking evaporation substantially \cite{Casadio:2001dc}. Recently, it has been discussed that PBHs production and their evaporation is also affected in a dimensionally reduced universe \cite{Tzikas:2018wzd}.}, PBHs evaporate on a time scale given by
\beq
t_{\rm ev}(M) \sim \frac{G^2 M^3}{\hbar\, c^4} \sim 10^{63} \l(\frac{M}{M_{\odot}}\r)^3 {\rm yr}.
\eeq
This implies that PBHs with mass $M \lesssim 10^{-18} M_{\odot}$ ($M \lesssim 10^{15}\, {\rm g}$) would be completely evaporated by today and thus can not contribute to the CDM today \cite{Carr:2009jm}. PBHs in the mass range $10^{-18} - 10^{-16} M_{\odot}$ would be evaporating at the present epoch and thus can induce an observable $\gamma$-ray background \cite{Laha:2019ssq}. In principle, this background could contribute to both the galactic and extragalactic  $\gamma$-ray background and the antiprotons or positrons in cosmic rays \cite{Carr:2009jm, Carr:2016hva}. It turns out that PBHs can not account for the total CDM if $M \lesssim 7 \times 10^{15}\, {\rm g}$. 
\item
$M \sim 10^{-16} - 10^{-11} M_{\odot}$:
There are mainly three classes of observational constraints available in this mass range. 
Firstly, compact objects such as PBHs can induce gravitational femtolensing of $\gamma$-ray bursts. The non-detection of any femtolensing signature in the Fermi Gamma-Ray Burst Monitor experiment led to strong constraints on the PBHs abundance in the mass range $10^{17} - 10^{20}\, {\rm g}$ \cite{Barnacka:2012bm}. Inclusion of the extended size of the source and wave optics effects, relaxes these constraints largely \cite{Katz:2018zrn}. Secondly, dark matter consisting of PBHs can cause the white dwarfs (WD) to explode as a supernova.  The shape of the observed distribution of WDs rules out PBHs within mass range $10^{19}-10^{24}\, {\rm g}$ \cite{Graham:2015apa}. These bounds have been obtained for a specific maximum mass and radius of WDs and are subject to astrophysical uncertainties. Lastly, if a PBH is captured by a neutron star, the star is accreted onto the PBH and gets destroyed in a relatively short time which constrains the PBHs mass fraction in the  range $10^{18}-10^{24}\, {\rm g}$ \cite{Capela:2013yf}. However, it suffers from astrophysical uncertainties in the CDM density inside globular clusters and neutron star properties such as mass, radius, lifetime and the velocity distribution.

Recent re-analysis indicates all these three classes of constraints to be ineffective and shows that this mass window ($M \sim 10^{-17} -  10^{-12} M_{\odot}$) essentially remains unconstrained for PBHs to constitute all dark matter \cite{Montero-Camacho:2019jte}. However, in the case of WDs, when PBHs pass through them without being gravitationally captured, it can trigger a supernovae, leaving observational imprints \cite{Graham:2015apa}. Therefore, we only consider the constraints arising from the WDs and not the other two, as shown on the right in Fig. \ref{sps}.

\item 
$M \sim 10^{-11} - 10^{-6} M_{\odot}$:
If PBHs dominate the galactic dark matter halo, one expects gravitational microlensing signatures of stars in the Magellanic cloud. However, such events are not detected in the haloes of the Milky Way (MW) and Andromeda galaxy (M31) by the Subaru Hyper Suprime-Cam. Simultaneous monitoring of millions of stars in M31 combined with long observations eliminate any possibility of PBHs abundance in the mass range $10^{22}-10^{27}\, {\rm g}$ although point source approximation leads to stronger bounds than finite source approximation in this case \cite{Niikura:2017zjd}.
\item 
$M \sim 10^{-7} - 15 M_{\odot}$:
Long monitoring of stars in the Magellanic clouds for microlensing events by EROS and OGLE surveys leads to a constraint on the mass range as $10^{26}-10^{34}\, {\rm g}$ at $2 \sigma$ and are also subject to various astrophysical uncertainties \cite{Tisserand:2006zx, Alcock:1998fx, Niikura:2019kqi}.
\item 
$M > 10M_{\odot }$:
Emission due to gas accretion onto PBHs leads to a considerable modification in the recombination history of the universe, thereby modifying the CMB temperature anisotropies and spectral distortion signatures. This results to the fact that PBHs with $M > 2 \times 10^{34} {\rm g}$ can at most contribute to a few percent of the total CDM in the universe \cite{Ricotti:2007au, Blum:2016cjs, Poulin:2017bwe}. However, these bounds assume that PBHs mass remains unchanged with time which is certainly not the case. Also, robust upper limit arises from the dynamics of ultra faint dwarf galaxies (UFD) observations \cite{Brandt:2016aco, Koushiappas:2017chw} and from the gravitational lensing of type Ia supernovae \cite{Zumalacarregui:2017qqd}.
\item
$M \sim 10^{2} - 10^{3} M_{\odot }$:
These constraints arise from a comprehensive analysis of high resolution and high redshift Lyman-$\alpha$ forest data. The fluctuations in the number density of PBHs induce an enhancement at small scales which departs from the standard CDM prediction. Using hydrodynamic simulations with different astrophysical parameters, one arrives at the mass range $10^{35}-10^{37}\, {\rm g}$ at $2 \sigma$ for a monochromatic PBH mass distribution which further become stronger for a non-monochromatic distribution \cite{Murgia:2019duy}.
 
\end{enumerate}

As mentioned earlier, all these constraints have been obtained by assuming a monochromatic mass spectrum for PBHs and in some cases, these constraints become tighter for the non-monochromatic mass spectrum \cite{Carr:2017jsz}. However, note that all these bounds involve various degrees of uncertainties such as uncertainties in the accretion process and its effect on the thermal history of the universe, properties of the MW halo in case of microlensing constraints, non-homogeneous spatial distribution of PBHs etc. and are very sensitive to assumptions about different astrophysical parameters. The current constraints for a monochromatic mass spectrum of PBHs are shown on the right in Figure \ref{sps}. It is evident from this figure that PBHs can indeed constitute the entire CDM around the asteroid-mass window.

PBHs can be produced by several mechanisms in the early universe. These include first-order phase transitions \cite{Jedamzik:1999am}, grand unified theories \cite{Khlopov:1980mg}, resonant reheating \cite{Suyama:2004mz}, tachyonic preheating \cite{Bassett:2000ha, Suyama:2006sr}, curvaton scenarios \cite{Kawasaki:2012wr, Kohri:2012yw, Bugaev:2013vba}, and inflationary scenarios \cite{Ivanov:1994pa, GarciaBellido:1996qt, Kawaguchi:2007fz, Kohri:2007qn, Drees:2011yz, Bugaev:2011wy, Erfani:2013iea, Clesse:2015wea, Erfani:2015rqv, Ezquiaga:2017fvi, Kannike:2017bxn, Hertzberg:2017dkh, Pi:2017gih, Cicoli:2018asa, Kamenshchik:2018sig, Dimopoulos:2019wew}. Recently, it has also been discussed if PBHs can be produced by a long range attractive fifth force stronger than the gravitational force in the early universe, mediated by a light scalar \cite{Amendola:2017xhl}. Among these various mechanisms, inflation, in particular, provides an ideal setting to produce PBHs in the very early universe. When small scale perturbations with large overdensities re-enter the expanding horizon during the radiation dominated epoch, they can collapse very quickly and form the PBHs. 
Their mass is typically given by the horizon mass at the time of re-entry of a given mode characterised by the wavenumber $k$ \cite{Carr:1975qj},
\be
M(k) \equiv \gamma M_H &=& \frac{4 \pi}{3} \frac{\gamma \rho}{H^{3}}\biggm \vert_{k=aH} \simeq 10^{18} \l(\frac{\gamma}{0.2}\r) \l(\frac{g_{\ast}}{106.75}\r)^{-\frac{1}{6}}\l(\frac{k}{7 \times 10^{13} \,{\rm Mpc^{-1}}}\r)^{-2}\! {\rm g}\,,
\ee
where $\gamma$ characterises the efficiency of the formation process ($\gamma \sim 0.2$) and $g_{\ast}$ is the effective number of the relativistic degrees of freedom at the formation epoch. This relation indicates that PBHs in wider mass ranges can be produced when very small scales re-enter the horizon during the radiation dominated epoch.
Due to Hawking evaporation, very light PBHs $(M \lesssim 10^{-18} M_{\odot})$ would be completely evaporated by today \cite{Carr:2009jm}. This leads to an interesting constraint on $k \lesssim k_{\ast} = 5 \times 10^{14} \, {\rm Mpc^{-1}}$ {\it i.e.} the lower bound on the physical length scales relevant for PBHs production such that they survive until today. The physical scales corresponding to $k \lesssim k_{\ast}$ are much smaller than the cosmological scales probed by the cosmic microwave background (CMB) anisotropies and large scale structure (LSS) surveys.  While the amplitude of the primordial  spectrum at the CMB scales is $P_{\zeta} \sim 10^{-9}$, it turns out that the spectrum must quickly rise to $\sim 10^{-2}$ at the relevant scales ($k \lesssim k_{\ast}$) in order to produce a sufficient abundance of PBHs such that their corresponding masses always satisfy the condition $M \gtrsim10^{-18} M_{\odot}$ so as to survive until today \cite{Harada:2013epa, Musco:2012au, Musco:2008hv, Musco:2004ak, Shibata:1999zs}. 

In general, not all the inflationary models can produce the relevant abundance of PBHs as it critically depends on the shape and various parameters of the potential. Recently, it has been pointed out that inflationary models with potentials having an inflection point are very useful in producing PBHs and thus polynomial potentials have often been used for this purpose \cite{Garcia-Bellido:2017mdw, Ballesteros:2017fsr}. 
However, simple polynomial potentials could be very steep at large field values and therefore, may not be able to produce a nearly scale invariant spectrum of primordial fluctuations at the CMB scales. One therefore needs to appropriately flatten out the potential at these scales such that the CMB observables are consistent with the Planck data.
In this paper, we study the abundance of PBHs generated in a single field inflection point model of inflation whose potential is motivated from the effective field theory of inflation and it is expanded upto the sextic power of the field $\phi$ with a cutoff scale $\Lambda$. Due to a flattening term in the potential, our effective potential actually behaves as a quadratic potential at both large and small values of $\phi$ which makes it renormalizable as well as consistent with the upper bound on the tensor-to-scalar ratio $r$ from Planck + BICEP2 \cite{Array:2015xqh, Ade:2018gkx}. While Planck data favours plateau models of inflation,  chaotic models can still be viable.

We study the background scalar field dynamics with this potential and find that it generically allows a dynamical phase wherein the fractional variation per e-fold of the first slow roll parameter becomes large and thus leads to a violation of the slow roll approximation. We implement a numerical scheme to evolve the background dynamics and linear perturbations and evaluate the spectrum at the end of inflation. We find that the slow roll spectrum computed at the horizon exit is very different from the spectrum evaluated numerically at the end of inflation. We then incorporate the Press-Schechter formalism in our code to compute the PBH mass fraction and find that PBHs can be produced for very different mass ranges for different choice of parameters of our model while being consistent with the CMB constraints on large scales. We further study the effects of a reheating phase after the end of inflation on the PBH mass fraction and find that an epoch of a matter dominated reheating can shift the PBHs mass fraction to a larger mass window as well as increase their fractional contribution to the total CDM for the case of a monochromatic mass fraction. Our results are broadly consistent with \cite{Cai:2018rqf} which are derived using the slow roll spectra for a different model.

The remainder of this paper is organised as follows: In the following section, we shall describe in detail the dynamics of the inflationary model with an inflection point that we explore and obtain the power spectra of curvature perturbations both using the slow roll formalism and an exact numerical computation. In Section \ref{pbh-mass-fraction}, we briefly discuss the Press-Schechter formalism to compute the PBHs mass fraction. In Section \ref{results}, we present our main results, highlighting the fact that PBHs can be produced in our scenario for very different mass ranges by varying different parameters. In order to achieve this, we also discuss a strategy to effectively scan the parameter space to arrive at the desired mass fraction. We also study the effects of reheating on the PBHs mass fraction and point out that the resulting effects could be very significant. Finally, we conclude and discuss the implications of our results in Section \ref{conclusions}. 
In appendices \ref{app-A} and \ref{app-B}, we discuss the background solutions and the uncertainties associated with the calculation of the PBHs mass fraction, respectively. 

Before we proceed further, a few words on the conventions and notations adopted in this paper are in order. We shall work in the natural units, $\hbar =c=1$, with reduced Planck mass $M_{\rm Pl}^2 = (8 \pi G)^{-1}$. Our metric signature is mostly plus with $(-, +, +, +)$. The background metric is described by the homogeneous, isotropic and spatially flat FLRW universe with a line element $ds^2 = -dt^2 +a^2(t) d{\bf x}^2 = a^2(\tau) (-d\tau^2 + d{\bf x}^2)$. The overdots and primes denote the derivatives with respect to the coordinate time $t$ and the conformal time $\tau$, respectively and the conformal time $\tau$ is defined as $d\tau =dt/a(t)$. The physical Hubble parameter is defined as $H \equiv {\dot a}/a$ while the conformal Hubble parameter is given by ${\cal H} \equiv {a'}/a$.


\section{An inflationary model with a polynomial potential}

In order to achieve the sufficient abundance of PBHs, an amplification of the power spectrum of the order $P_{\zeta} \sim 10^{-2}$ at the appropriate scales is required. Since $P_{\zeta} \sim H^2/\epsilon$ , this feature can roughly be achieved in any single field inflationary model wherein the first slow roll parameter $\epsilon$ becomes very small at those scales. Although $\epsilon \ll 1$, its fractional variation per e-fold becomes large and thus leads to a violation of the slow roll approximation. Hence, it turns out that a necessary condition for the formation of PBHs is ${\cal O} (1)$ violation of the slow roll conditions. Since this violation is large enough, one can not {\it just} use the slow roll approximation to compute the power spectrum \cite{Chongchitnan:2006wx, Germani:2017bcs, Motohashi:2017kbs}. Therefore, in all such models, the power spectrum should be computed exactly by adopting a suitable numerical scheme.


\subsection{An effective scenario with an inflection point}

\begin{figure}[t]
\begin{center}
\includegraphics[width=10cm]{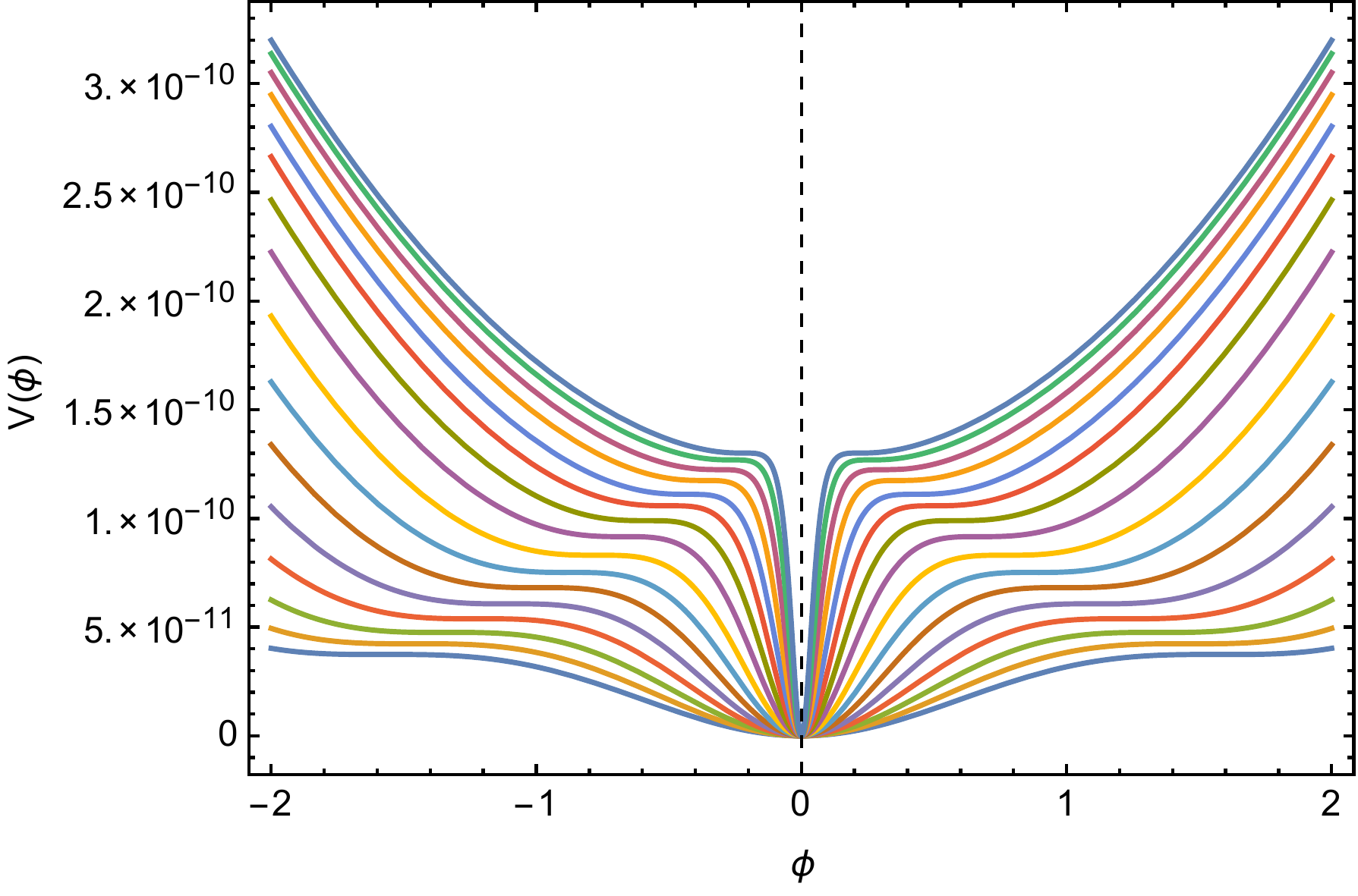} 
\caption{The scalar potential of our inflationary model is plotted for different choices of parameters corresponding to different values of quasi-inflection points. Depending on the choices of these parameters, an inflection point may or may not exist in the potential.  
In general, in such potentials, the large field region corresponds to the observed CMB scales, the plateau region leads to PBHs formation while the true minima corresponds to the reheating phase after the end of inflation. Note that, our potential effectively behaves as $\sim \phi^2$ at large values of $\phi$ corresponding to the CMB scales and is symmetric under $\phi \to -\phi$, unlike other examples discussed in the literature.}
\label{potential-plot}
\end{center}
\end{figure}

Besides a class of monomial potentials, polynomial potentials have also been used to drive inflation as well as to explain possible anomalies in the CMB such as the power suppression at low multipoles \cite{Jain:2008dw, Jain:2009pm}. Such a potential typically allows the existence of an inflection point (a plateau region) which leads to a phase to slow roll violation (due to ultra slow roll evolution) around that region. So far, polynomial potentials have been used in various contexts such as explaining large scale suppression in the CMB, producing sufficient abundance of PBHs and inducing an observably large tensor-to-scalar ratio \cite{BenDayan:2009kv, Hotchkiss:2011gz, Wolfson:2019rwd}. A polynomial potential can be motivated from the framework of an effective field theory (EFT) with a cutoff scale $\Lambda$ wherein the effective potential is generally given by \cite{Enqvist:2003gh, Burgess:2009ea, Marchesano:2014mla}
\be
\tilde V_{\rm eff} (\phi) &=& \sum_{n} \frac{b_{n}}{n\,!} \l(\frac{\phi}{\Lambda}\r)^n = \tilde V_{0} \l[1+c_{1}\frac{\phi}{\Lambda} + \frac{c_{2}}{2!}\l(\frac{\phi}{\Lambda}\r)^2+ \frac{c_{3}}{3!}\l(\frac{\phi}{\Lambda}\r)^3+ .... \r],
\ee
where $V_{0}$ is an overall factor, often referred to as the energy scale of the inflaton potential and $c_i$ are constants. In principle, the effective potential consists of all the higher order terms but in practice, one truncates the potential at some order to achieve the desired inflationary dynamics. Moreover, one can also impose certain symmetries such that the potential remains unchanged under $\phi \to -\phi$ symmetry while the constant term can be absorbed in the redefinition of the potential. By truncating the effective potential to the sextic order and ignoring the constant term, we can write the potential as \cite{Burgess:2009ea, Marchesano:2014mla}
\beq
\tilde V_{\rm eff} (\phi) = \tilde V_{0} \l[\frac{c_{2}}{2!}\l(\frac{\phi}{\Lambda}\r)^2+ \frac{c_{4}}{4!}\l(\frac{\phi}{\Lambda}\r)^4+ \frac{c_{6}}{6!}\l(\frac{\phi}{\Lambda}\r)^6 \r].
\eeq
This potential generically allows the existence of inflection points for different choices of the constants $c_i$ but it turns out to be non-renormalizable. If inflation has to take place at large values of $\phi$, one can show that inflationary observables will be in strong disagreement with the CMB as the potential is extremely steep at large values of $\phi$. Therefore, one must flatten the potential enough to get the desired inflationary dynamics. This flattening can be achieved by introducing an appropriate factor such that the potential can now be written as
\beq
V (\phi) = \frac{\tilde V_{\rm eff} (\phi)}{(1+\xi \phi^2)^2},
\eeq
where $\xi$ is also a constant. This potential can be recast in the form\footnote{Note that this potential that provides us the desired dynamics in our scenario can be considered as a combination of the motivation from the EFT and the requirement for the needed flatness at the CMB scales which also makes the potential renormalizable at large field values.}
\beq
V (x) = V_{0}\, \frac{ax^2+bx^4+cx^6}{(1+dx^2)^2}, \label{potential-exp}
\eeq
where $x=\phi/v$, $V_{0}=\tilde V_{0} (v/\Lambda)^2$, $a=c_2/2!$, $b=c_4/4!  (v/\Lambda)^2$, $c=c_6/6!  (v/\Lambda)^4$ and $d=\xi v^2$. The parameter $v$ is just a constant scaling factor. This is the potential of our scenario we shall work with in this paper. Note that, this potential behaves as a quadratic potential {\it i.e.} $V(\phi) \sim \phi^2$ for both the large and small field values (far away from the plateau region). The flattening of the potential helps in two ways: first, it makes the potential renormalizable at large field values and second, it flattens the potential sufficiently for large sales. 
Such a behaviour of the potential leads to a nearly scale invariant power spectrum of curvature perturbations which is consistent with the CMB observations on large scales. As we shall discuss later, this potential also dynamically leads to a phase of slow roll violation (or ultra slow roll) close to the plateau region and finally reheats the universe when the scalar field rolls down to the true minima of the potential. The asymptotic behaviour of our potential is very different from the flattened quartic polynomial potential which has recently been discussed in the literature \cite{Garcia-Bellido:2017mdw, Ballesteros:2017fsr, Hertzberg:2017dkh}. While our potential being roughly quadratic at large scales leads to an observably large tensor-to-scalar ratio $r$, their potential being nearly flat on large scale induces a very small $r$. Moreover, the quartic polynomial potential does not possess an inversion symmetry {\it i.e.} the potential is not symmetric under $\phi \to -\phi$ and thus, one has to necessarily start the dynamics from positive values of the field $\phi$. This restriction is relaxed in our scenario as our potential has an inversion symmetry and therefore, one can start from both the positive or negative values of $\phi$. The inflaton potential of our scenario is shown in Figure \ref{potential-plot} for different choices of parameters corresponding to different values of quasi-inflection points which are appropriate to obtain the desired inflationary dynamics. 

Before we proceed further and discuss the details of the presence of inflection points in our potential, let us briefly comment on the quartic polynomial potential proposed in \cite{Garcia-Bellido:2017mdw}
\begin{equation}
U(x)=U_0 \frac{a x^2+b x^3+c x^4}{\left(1+d x^2\right)^2}.   \label{pot-234}
\end{equation}
The denominator provides appropriate flatness of the potential and could possibly be motivated from a non minimal coupling term in the Lagrangian. However, this potential behaves as a constant potential at large values of $x$ (or $\phi$) while behaves as a quadratic potential at small values of $x$.  For the case of our potential given in \eqref{potential-exp}, we find that in the limit of $x \gg 1$, the potential reduces to
\beq
V(x)\simeq  \frac{V_0 c}{d^2} x^2,	\label{pot-ls}
\eeq
while in the small $x$ limit {\it i.e.} $x \ll 1$
\beq
V(x)\simeq  V_0 a\, x^2.	\label{pot-ss}
\eeq
Therefore, in both the limits, our potential reduces to a quadratic potential which is renormalizable and it is very different from the earlier potential. This is one of the key differences between the two cases. Since the CMB scales correspond to the large field part of the potential, as we shall discuss in Section \ref{results}, our scenario will lead to a rather large tensor-to-scalar ratio $r$ on such scales as compared to the earlier scenario. Future observations of the tensor-to-scalar ratio can strongly constrain our scenario.

We would now like to understand the details of various inflection points that originate in our potential. A necessary condition for the existence of an inflection point is that the lowest order (above the second) non-zero derivative to be of odd order. We start with the first and second derivatives of the potential given in \eqref{potential-exp} and set them to zero, leaving the third derivative non-zero. This leads to the following two equations given by
\begin{align}
a +(2 b- a d) x^2+3 c x^4+ c d x^6 & =0,			\label{p-eq-1}\\
a+ 2(3 b - 4 a d) x^2 +3(a d^2-2 b d+5 c )x^4+4 c d x^6 + c d^2 x^8& =0.		\label{p-eq-2}
\end{align}
In this case, we get two equality conditions and one inequality condition, but have five unknowns $a,b,c,d$ and $x$ to find. 
We shall try to arrive at conditions using these two equations for the existence of inflection point(s) and refer to them as the `inflection point conditions' which are given in rather simpler form as
\beq
a=\frac{b x^2 \left(2 d x^2+3\right)}{(d^2 x^4-3)} \qquad {\rm and} \qquad c=\frac{b}{x^2(d^2 x^4-3)}.								
\label{cond1}
\eeq
First, by looking at the potential in \eqref{potential-exp}, we can impose $d>0$ to avoid any singularities in the potential. 
Also, if the potential should be monotonously increasing for both large and small field values, we find, from \eqref{pot-ls} and \eqref{pot-ss}, that  $a>0$ and $c>0$.
If we limit ourselves in the positive field regime of the potential {\it i.e.} $x>0$, we notice that in order  to satisfy \eqref{cond1} together with $a,c >0$, we arrive at the following two conditions on $b$.
\begin{description}[font=$\bullet$~\normalfont\scshape\color{black!50!black}]
	\item [$ x>(3/d^2)^{1/4}$] leads to positive denominators and thus $b>0$.
	\item [$ x<(3/d^2)^{1/4}$]  leads to negative denominators and thus $b<0$.
\end{description}
Therefore, in this model, we get two different kinds of inflection points for two choices of $b$, $b>0$ and $b<0$ and they are separated by $ x \sim (3/d^2)^{1/4}$. Since it is not possible to explore the details of inflection points for this model analytically, we shall resort to a numerical scan of the parameter space to obtain real inflection points. In fact, we have noticed that if we first fix the value of the inflection point $x_0$ and the values of $b$ and $d$, one can then determine $a$ and $c$ using \eqref{cond1}.  Although the potential has four independent parameters, one only needs to fix two of them and the rest will be determined by the inflection point conditions. We shall discuss our numerical strategy to scan the parameter space in Section \ref{results}.

\subsection{Background evolution and the enhancement of curvature perturbations}
\label{bgnd}
\begin{figure}[!t]
\begin{center}
\includegraphics[width=10cm]{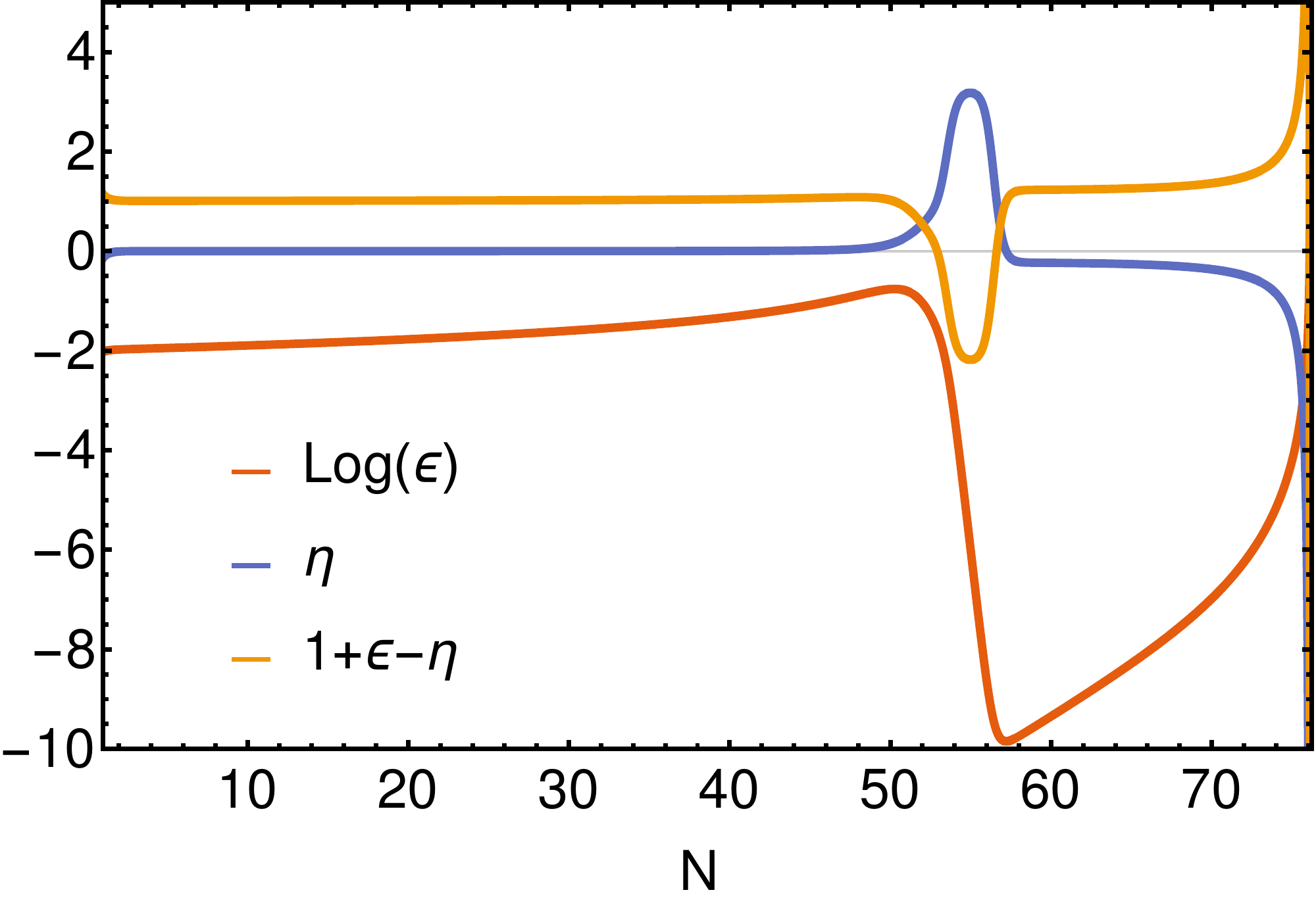}
\caption{The evolution of the slow roll parameters $\epsilon, \eta$ and the quantity $(1+\epsilon -\eta)$ as a function of $N$ around the ultra slow-roll regime for our potential given in \eqref{potential-exp}. Since $\epsilon$ becomes very small during the ultra slow-roll regime, we have chosen to show the behaviour of ${\rm Log}\, (\epsilon)$ so that its evolution during the entire e-folds domain is evident. Although $\epsilon$ remains much smaller than unity, its fractional variation per e-fold becomes significantly large, leading to a transient violation of slow roll. Note that, although $\epsilon$ increases quite rapidly after around $N \sim 58$, it still remains smaller than unity and hence, inflation is never interrupted in this scenario during the transient phase. }
\label{3_back}
\end{center}
\end{figure}

The presence of an inflection point in the potential leads to a phase of ultra slow roll in the background evolution while the dynamics away from the inflection point is slow roll. Given a potential, the background dynamics can be studied using the Friedmann and the Klein-Gordon equations for the inflaton $\phi$. All the equations and their solutions in various regimes are summarised in appendix \ref{app-A}. While $\phi$ monotonically decreases with the number of e-folds $N$ during slow roll, the inflaton velocity gets exponentially suppressed during the ultra slow roll phase which leads to $\epsilon \sim {\rm exp}\, [-6N]$ and $\eta \sim 3$, as evident in Figure \ref{3_back}.
As we shall discuss below, this dynamical evolution during the ultra slow roll phase leads to an exponential growth of the power spectrum around the scales relevant for the PBHs formation.

As mentioned earlier, a necessary condition for the production of PBHs from inflation leads to the fact that the
spectrum of primordial curvature perturbations has to increase by a factor of $\sim 10^7$ or so in its amplitude at scales much smaller than the observable CMB scales. Such an enhancement can induce large matter density fluctuations at the horizon re-entry of these scales which can collapse to form PBHs. Let us briefly review the condition under which such an enhancement can happen \cite{Leach:2000yw, Leach:2001zf, Jain:2007au, Ozsoy:2018flq}. We start with the Fourier mode equation for the  comoving curvature perturbation ${\cal R}_k$ in single field inflation which follows from the well known Mukhanov-Sasaki equation 
\beq
{\cal R}_k''+2\left(\frac{z'}{z}\right){\cal R}_k'+k^2{\cal R}_k=0 . \label{MS-1}
\eeq
The comoving curvature perturbation ${\cal R}_k$ matches with the curvature perturbation on the constant density hypersurface $\zeta_k$ in the super-horizon limit. In the rest of this paper, we shall use these two variables interchangeably to describe the scalar power spectrum. The `pump field' $z$ is defined as $z=a \dot{\phi}/{H}$ and the friction term $z'/z$ is given by
\beq
\frac{z'}{z}=aH (1+\epsilon  -\eta). \label{zprimez}
\eeq
During slow roll inflation, $\epsilon, |\eta| \ll 1$ and thus, $z \sim a$. In the super-horizon limit $k \ll aH$, a very general solution to \eqref{MS-1} can be expressed as \cite{Leach:2001zf}
\beq
{\cal R}_k(\tau) \simeq C_1  + C_2 \int\frac{d\tau}{z^2}.\label{MS-3}
\eeq
In slow roll inflation, it is evident to note that the second term decays rapidly as $a^{-3}$ outside the horizon and therefore, the curvature perturbation ${\cal R}_k$ is conserved in time at super-horizon scales for each wavenumber $k$ and its amplitude is determined by the constant $C_1$. Since the decaying mode dies completely in this limit, one can safely compute the power spectrum of curvature perturbations at the horizon exit in slow roll by fixing the initial conditions for each Fourier mode using the Bunch-Davies vacuum at sub-horizon scales. 

However, a complete different situation arises if the friction term $z'/z$ transiently changes sign and becomes a `driving term' during an epoch for different modes right after they cross the horizon. This is equivalent to the following condition
\beq
(1+\epsilon  -\eta) < 0\,.
\label{cond-usr}
\eeq
Since $\epsilon >0$ by definition, this condition can {\it only} be achieved if $\eta > 1$ during some epoch. This can be satisfied if the universe goes through a dynamical phase during which the slow-roll approximation breaks down. In particular, the ultra slow-roll regime that we mentioned in the introduction, corresponds to a dynamical phase during which $\eta \ge 3$.
During this regime, the condition \eqref{cond-usr} is satisfied and hence $z$ now decreases with time instead of increasing. This implies that the otherwise decaying mode appearing in \eqref{MS-3} now becomes a growing mode in this regime and its contribution to the curvature perturbation ${\cal R}_k$ can no longer be neglected. This transient growth of the decaying mode can be used to enhance the primordial spectrum of curvature perturbations for a short range of scales and thus, to produce PBHs on these scales. However, one has to fine tune a given inflationary scenario to obtain this transient departure from slow roll such that PBHs are abundantly produced on appropriate scales so as to contribute to a larger fraction to the CDM. 

In general, in any potential with a plateau region as in our scenario, this transient phase with a departure from slow roll is very likely to be present. In our model, we find that this ultra slow roll regime is present for a few e-folds corresponding to the scales appropriate for PBHs formation, as shown in Figure \ref{3_back}.  It is the dynamics of this transient regime that is responsible for the required exponential growth of ${\cal R}_k$ resulting in a suitable bump in the $P_{\zeta}(k)$ as high as $\sim 10^{-2}$ needed for the PBHs formation.

\section{Primordial black holes mass fraction and the collapse criteria }
\label{pbh-mass-fraction}

In this section, we shall quickly discuss the basic formalism to calculate the abundance of PBHs from a primordial power spectrum obtained at the end of inflation. In appendix \ref{app-B}, we shall also point out various uncertainties in the estimation of the final mass fraction which are associated with the underlying collapse criteria, the choice of the window function and the value of the critical density contrast. As we shall discuss later, it turns out that the mass fraction is exponentially sensitive to the value of the critical density contrast and can significantly change the predictions of a model for the same primordial power spectrum.

The mass fraction $\beta (M)$ which specifies the fraction of the energy density of the universe populated by PBHs formed with a mass $M$ is an important quantity, defined by
\beq
\beta(M)\equiv \frac{\rho_{_{\rm PBH}}(M)}{\rho_{\rm tot}}.
\eeq
The mass fraction is a very important quantity in this context which is typically calculated at the time of the PBH formation $t_{f}$ but is generally translated to the epoch of radiation-matter equality $t_{\rm eq}$ to compare it with the CDM fraction in the universe at that epoch. Since PBHs behave as matter, $\rho_{_{\rm PBH}} \sim \rho_{\rm m} \sim a^{-3}$ and since they are formed during radiation domination, $\rho_{\rm tot} \sim \rho_{\rm rad} \sim a^{-4}$, this implies that $\beta (M) \sim a$ {\it i.e.} $\beta (M)$ grows with the scale factor until the radiation-matter equality. This relation can be used to arrive at the following expression relating the mass fraction at the epochs of formation and the radiation-matter equality as
\beq
\beta_{\rm eq}(M) \simeq \beta_{f}(M) \l(\frac{a_{\rm eq}}{a_{f}}\r), \label{betarelation}
\eeq
where $a_f$ is the scale factor at $t=t_f$ i.e at the time of formation of PBHs. Since one assumes that PBHs are formed immediately by the collapse of a given scale at its horizon re-entry, $a_f$ can actually be calculated using the relation $k=aH$. 
A very useful quantity is the total abundance of PBHs which is given by \cite{Inomata:2017okj}
\beq
\Omega_{\rm PBH} = \int \frac{dM}{M} \Omega_{\rm PBH} (M),
\eeq
where
\be
f_{\rm PBH} (M) &\equiv& \frac{\Omega_{\rm PBH}(M)}{\Omega_{\rm DM}} =\frac{\beta(M)}{8 \times10^{-16}} \l(\frac{\gamma}{0.2}\r)^{3/2}  \l(\frac{g_{\ast}}{106.75}\r)^{-1/4}\l(\frac{M}{10^{18} \,{\rm g}}\r)^{-1/2} ,
\ee
and the total CDM fraction is constrained at equality to be $\Omega_{\rm DM} \simeq 0.42$ \cite{Ade:2015xua}. In a given inflationary model, the aim is to obtain the largest allowed value of $f_{\rm PBH}$ in a given mass range. This not only suffers from the fine-tuning of various parameters of the model but also turns out to be extremely sensitive to the peak in the primordial spectrum $P_{\cal R}$ or the choice of $\delta_c$ which we fix as $\delta_c=0.414$ as discussed in appendix \ref{app-B}. In this paper we shall essentially limit ourselves to the mass fraction in a very narrow mass range (a monochromatic mass fraction) and shall not consider the cases wherein PBHs are produced in a continuous broad mass range or discretely at different mass ranges. In such cases, the mass fraction $\beta(M)$ is conventionally described  by a mass function $\psi(M)$. 

We now employ the Press-Schechter formalism to compute the PBH mass fraction. We start with the simple relation between the comoving curvature perturbation ${\cal R}_{k}$ and the density contrast $\delta(t,k)$ which is given by \cite{Liddle:2000cg, Green:2004wb, Ballesteros:2018wlw}
\begin{equation}
\delta(k,t) \simeq \frac{2(w +1)}{(3 w +5)}\l(\frac{ k}{aH}\r)^2 {\cal R}_{k}.     \label{delta1}
\end{equation}
At the horizon re-entry of a given mode in the radiation dominated epoch with $w=1/3$, one finds that $\delta = (4/9) {\cal R}$. Using this, we can obtain the following relation between the power spectra of $\delta$ and ${\cal R}_{k}$ 
\begin{equation}
P_{\delta}(k,t) \simeq  \frac{4(w +1)^2}{(3 w +5)^2}\l(\frac{k}{aH}\r)^4 P_{\cal R}(k).    \label{pdelta}
\end{equation}
Now, the variance of the density contrast $ \sigma_\delta^2 $ at a comoving scale $R$, course grained using a Gaussian window function described in \eqref{window}  can be expressed as
\begin{equation}
\sigma_\delta^2 (t,R)=\int \frac{dk}{k}P_{\delta}(k,t) W^2(k,R).              \label{vdelta}
\end{equation}
For PBHs which are formed due to the collapse of a comoving wavenumber $k$ just after the horizon re-entry during the radiation epoch, we can absorb the time dependence in the comoving smoothing scale, $R=(aH)^{-1}$. This leads to 
\begin{equation}
\sigma_\delta^2 (R) \simeq \frac{16}{81} \int \frac{dk}{k}(kR)^4 P_{\cal R}(k)  W^2(k,R).       \label{vdelta2}
\end{equation}
In the Press-Schechter formalism of gravitational collapse \cite{Press:1973iz}, the mass fraction $\beta(M)$ in PBHs of mass $M$, is given by the probability that the overdensity $\delta$ is above a certain threshold value $\delta_c$ for collapse. Assuming $\delta$ is a Gaussian random variable with mass (or scale) dependent variance, the mass fraction $\beta(M)$ at the time of formation is then given by\footnote{Note that, one can include here the famous ``fudge factor" of $2$ of the Press-Schechter formalism as it is done conventionally.}
\begin{equation}
\beta_{f}(M)= \frac{1}{\sqrt{2 \pi \sigma_\delta^2 (M(R))}}\int_{\delta_c}^\infty d\delta\, {\rm exp}\l(-\frac{\delta^2}{2 \sigma_\delta^2 (M(R))}\r)
=\frac{1}{2}{\rm erfc}\l(\frac{\delta_c}{\sqrt{2}\, \sigma_\delta (M(R))}\r),     \label{ps1}
\end{equation}
where ${\rm erfc(x)}$ is the complementary error function. Often, one can use the fact that it is the fluctuations in the upper tail of the distribution that form the PBHs and therefore, the complementary error function in the above expression can be approximated by an exponential function so that the mass fraction is given by \cite{Hertzberg:2017dkh}
\beq
\beta_{f}(M) \simeq \sqrt{\frac{1}{2 \pi}}\frac{\sigma_\delta (M(R))}{\delta_c} {\rm exp}\l(-\frac{\delta_c^2}{2 \sigma_\delta^2 (M(R))}\r).
\label{ps1approx}
\eeq
In order to proceed further,  we first need to relate the comoving smoothing scale $R$ to the PBH mass at formation. Since every smoothing scale $R$ corresponds to a formed PBH of comoving radius $R$,  the formation mass is simply given by 
\beq
M(R)= \frac{4 \pi}{3}\gamma \rho\, (aR)^3 ,       \label{mpbh1}
\eeq
where $\gamma \sim 0.2$ is the efficiency factor as mentioned in the introduction and $R=(aH)^{-1}$ at the horizon re-entry of that scale. As PBHs in our scenario are formed when the respective scales re-enter the horizon during radiation domination, we have $a(t) \sim t^{1/2}$ so $aH \sim 1/a$ and thus $R \sim a$. Upon using this, we can find the following useful relation, ${R_{f}}/{R_{eq}}={a_{f}}/{a_{eq}}$ where $ R_{eq}^{-1} \sim k_{eq} = 0.07\, \Omega_m h^2 \,{\rm Mpc}^{-1}$ and $a_{eq}^{-1}=24000 \, \Omega_m h^2$. 
Using this together with  \eqref{mpbh1} and $H_{f}=(a_{f} R_{f})^{-1} $, we can now find the PBHs mass at formation as
\beq
M(R_{f}) = 4\pi\gamma M_{\rm Pl}^2 \l(\frac{a_{eq}}{R_{eq}}\r)  R_{f}^2\,. \label{mpbhf}
\eeq
This equation can be used to calculate the mass fraction $\beta_f(M)$ from \eqref{ps1} or \eqref{ps1approx}.  Once $\beta_f(M)$ is calculated, we can readily use \eqref{betarelation} to calculate the mass fraction at equality which is given by \cite{Cai:2018rqf}
\beq
\beta_{eq}(M) \simeq  \beta_{f}(M) \l(\frac{a_{eq}}{a_{f}}\r) = \beta_{f}(M) \l(\frac{R_{eq}}{R_{f}}\r). \label{beta2}
\eeq
Assuming a monochromatic mass spectrum, the PBHs fraction in the form of dark matter can be expressed as \cite{Cai:2018rqf}
\beq
f_{\rm PBH} (M) = \frac{\Omega_{\rm PBH}(M)}{\Omega_{\rm DM}} \approx \frac{\Omega_{\rm PBH}^{eq}(M)}{0.42},
\eeq
where $\Omega_{\rm PBH}^{eq}(M) \sim \beta_{eq}(M_{max})$ and $M_{max} $ is the PBH mass which contributes the maximum mass fraction at the radiation-matter equality.


\section{Numerical framework and results}
\label{results}

\subsection{Limitations of the slow roll approximation and the exact power spectra}

Let us start with the initial conditions $\phi_{i}$, $d\phi_i/d{N}$ and $H_i$ we need for the background evolution. 
The initial value of the inflaton field $\phi_{i}$ is important for two reasons: first, a minimum value of $\phi_{i}$ ensures that the field acquires enough momentum to classically\footnote{Since the inflaton field in the ultra slow roll regime undergoes a phase of strong deceleration, the quantum diffusion of the inflation becomes very relevant around the inflection point. We shall discuss its implications for our model elsewhere \cite{future-qf-pbh}.} cross over the inflection point within a finite number of e-folds $N$ and second, the slowing down of the inflation near the inflection point depends, exclusively on the choice of other parameters in the potential and the nature of the inflection point, but it also largely depends on $\phi_{i}$.  Once the value of $\phi_{i}$ is appropriately chosen, the initial values of $d\phi/d{N}$ and Hubble parameter $H_i$ can be obtained from the slow roll conditions and the Friedmann equations, respectively.

After the background dynamics is completely described, we notice that for a particular set of parameters which satisfy the inflection point conditions, the duration of inflation in terms of $N$ can be a subject of fine tuning in terms of $\phi_{i}$ or the ratios of parameters in the potential. Since any change in the value of $\phi_{i}$ affects the mass range of PBHs in our model, we often need a different set of parameters to obtain the right mass PBHs with inflection point conditions unharmed.  Since it is very impractical to change all the parameters simultaneously, we find that if we rescale the field as  $x=\phi/v$, we can overcome this issue saliently. 
It is important to note that, any non-zero constant value of $v$ would satisfy the inflection point conditions. From eq. \eqref{cond1}, if we fix all the coefficients $a,b,c,d$ of our potential,  different values of $v$ will lead to drastically different values of $N$ while for fixed number of e-folds, the location of the bump in the scalar power spectra required for PBHs formation, will roughly depend on the choice of $\phi_{i}$ and the location of the inflection point $x_0$. This freedom allows us to work with fixed values of $x_0$ and $\phi _{i}$ and explore the parameter space in terms of $a,b,c,d$ by changing $v$ only. With this, we also find that the duration of the ultra slow roll phase and the value of $N$ depend strongly on $v$ and one can get the desired value of $N$ by changing $v$ only and not any other parameters. From a numerical perspective, we can get a precise value of $v$ corresponding to a fixed number of e-folds. While computing the scalar and tensor power spectra, we always ensure that the universe always undergoes the minimal required number of e-folds. 

For exact numerical analysis, we also need the initial value of the scale factor and it must be chosen appropriately so as to lead to the correct length scale $k_{eq}$ of the radiation-matter equality. It is important to note that $k_{eq}$ is completely fixed by the energy densities of radiation and matter today. The choice of initial value of the scale factor depends on the number of e-folds during inflation and initial conditions of the radiation dominated phase which eventually depends on the inflationary dynamics. The ratio $a_i/a_0$ can be written as 
\be
\frac{a_{i} }{a_{0}} = \frac{a_{i} }{a_{e}} \frac{a_{e} }{a_{0} } \qquad {\rm and} \qquad \frac{a_{i} }{a_{e} } = e^{-N},
\label{a-eq}
\ee
where $a_{i}$, $a_{e}$ and $a_{0}$ are the scale factors at the beginning, end of inflation and at the present time, respectively and $N$ is the total number of e-folds during inflation. In the case of an instantaneous reheating phase wherein the universe becomes radiation dominated right after the end of inflation, one can write
\beq
\frac{a_{e} }{a_{0}} =\frac{a_{r} }{a_{0}}.
\eeq
However, if the reheating is not instantaneous, the above equation will be modified and we shall discuss the effects of such an epoch in the following section. Now that aim is to determine the ratio $a_{r}/a_{0}$.  From the equation of entropy conservation for relativistic particles, we can write
\beq
S=\frac{U+PV}{T}=\frac{\rho_r V+\frac{1}{3}\rho_r V}{T}=\frac{4}{3} \frac{\rho_r V}{T}={\rm constant},
\eeq
where $V\propto a(t)^3$ and $\rho_r$ is the radiation energy density at any epoch and is given by
\beq
\rho_r=\frac{\pi^2}{30} g_*(T)\,  T^4  ,
\label{eq_entrop}
\eeq
which leads to $T \propto g_*^{-1/3}/a$. Here,  $g_* (T)$ is the total number of effective degrees of freedom taking into account all the relativistic particles, $T$ is the photon temperature and both of them evolve with time. Now, using the entropy conservation at the beginning of radiation dominated phase and at the present epoch, we find 
\beq
\frac{a_{r}}{a_{0}}=\l(\frac{g_{*,0}}{g_{*,r}}\r)^{1/3}\frac{T_{\gamma}}{T_{r}},
\label{a-r0}
\eeq
where $g_{*,0} =3.36$, $g_{*,r}=g_{*,e}=106.75$ and $T_{\gamma}=2.725$ K. Upon using \eqref{a-eq} and \eqref{a-r0}, we can now write
\beq
\frac{a_{i} }{a_{0}} = \l(\frac{g_{*,0}}{g_{*,r}}\r)^{1/3}\frac{T_{\gamma}}{T_{r}} \,e^{-N},
\label{a-i0}
\eeq
which will provide the value of $a_i$ for our numerical calculations. At the end of inflation, the energy density can be obtained using Friedmann equations as $\rho_{e}=3 H_{e}^2 M_{\rm Pl}^2$. 
Using this and $\rho_{e}=\frac{\pi^2}{30} g_{*,e} T_{e}^4$, one can now find the temperature at the end of inflation as
\beq
T_e = \l(\frac{90}{\pi^2 g_{*,e}}\r)^{1/4} \l(H_{e} M_{\rm Pl}\r)^{1/2},
\eeq
which should also serve as the reheating temperature for the case of instantaneous reheating. In our numerical approach, each time a different initial value of the scale factor is chosen, it leads to different physical length scales and different conditions for the normalisation of the scalar spectrum at pivot scale, and also different initial conditions for the radiation dominated phase in terms of $H_{e}$ and $T_{e}$.  So while satisfying all these equations for consistency, only an adaptive numerical search of the parameter space can suggest an appropriate value of the initial scale factor which must be consistent with the CMB normalisation.


\subsection{Primordial power spectra and the PBH mass fraction}

As we have discussed earlier, a plateau region corresponding to an inflection point in the potential will generally lead to an ultra slow roll evolution which is required for the growth of the scalar power spectra at scales relevant for the PBHs formation. An exact inflection point that demands, exact zero values for the first and second derivatives of potential, does not always offer a very appropriate PBH mass fraction. Moreover, a quasi-inflection point with a very small positive value (very close to zero) of the first derivative of the potential leads to a slightly lower bump in the power spectrum than the exact inflection point case whereas a quasi-inflection point with a small negative value of the first derivative leads to a higher bump in the power spectra. The aim of our numerical approach is to choose a deviation from an exact inflection point in such a way that the contribution to the PBH mass fraction is maximum allowed in the relevant mass range.

In general, the primordial power spectrum of curvature and tensor perturbations are defined as
\beq
P_{\cal R}(k) = \frac{k^3}{2 \pi^2} \lvert {\cal R}_{k}\rvert^2 , \qquad P_{h}(k) = 2 \frac{k^3}{2 \pi^2} \lvert h_{k}\rvert^2
\eeq
and the Bunch-Davies vacuum initial conditions on $ {\cal R}_{k}$ and $h_k$ are imposed in the sub-horizon regime $k \gg aH$ which are given by
\beq
{\cal R}_k  = \frac{1}{\sqrt{2k}} \frac{e^{-ik\tau}}{z}, \qquad h_k  = \frac{1}{\sqrt{2k}} \frac{e^{-ik\tau}}{a}. 
\eeq
The tensor perturbations $h_k$ follow the same equation as \eqref{MS-1} but with $z \to a$. The spectra are generally calculated in the super-horizon limit wherein $k \ll aH$ while ensuring that the universe always inflates for the minimal required number of e-folds. In order to produce a large enough abundance of PBHs, the power spectrum must grow exponentially from an amplitude of $\sim 10^{-9}$ at CMB scales to $\sim 10^{-2}$ at the PBH scales and this is controlled by the smallness and the rate of change of $\epsilon$. As has been discussed in earlier works, this kind of inflationary dynamics around a local minima (or an inflection point) can not be correctly captured by the slow roll approximation \cite{Ballesteros:2017fsr}. Since the PBH mass fraction also depends exponentially on the density contrast $\delta$, an exact numerical integration of the mode equation for curvature perturbations is required to calculate the power spectrum which also justifies our approach of calculating all the relevant quantities exactly by adapting a suitable numerical scheme without relying on any approximations. 

\begin{figure}[t]
\centering
\includegraphics[width=7.5cm, height=5.4cm]{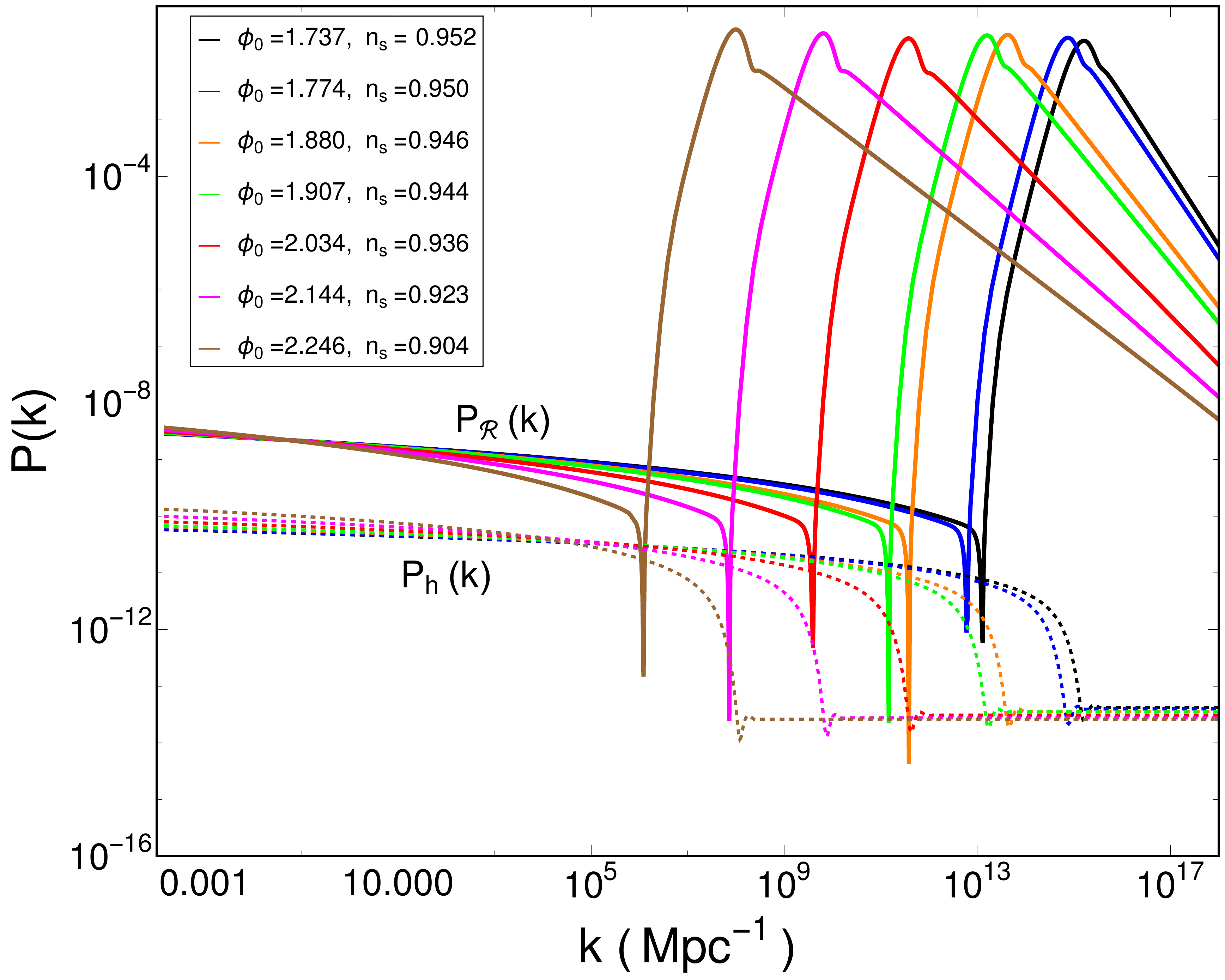}
\includegraphics[width=7.8cm, height=5.4cm]{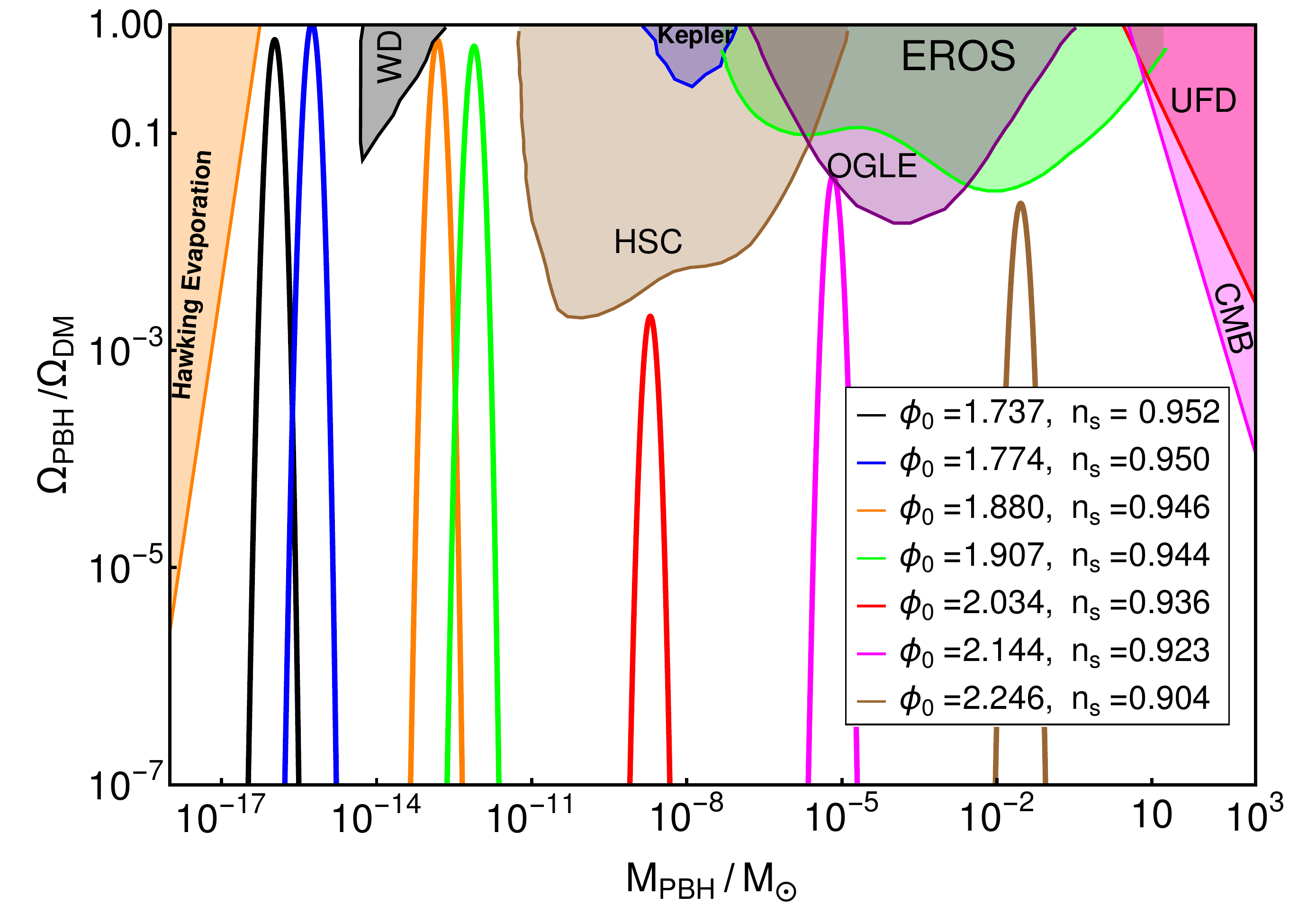}
\caption{On the left, the scalar power spectra $P_{\cal R}$ (solid lines) are shown for our model corresponding to different values of the inflection point as displayed in the inset. All the spectra show a similar exponential growth around the scales associated with the ultra slow roll regime. 
We also compute the tensor power spectra (dashed lines) for all the different cases. While the scalar power spectra show a tremendous rise for all the different cases, the tensor spectra result in a suppression on the smaller scales. As the bump in the scalar spectra shifts to the left, the tension of the spectral index $n_s$ with the best fit value from Planck increases.
On the right, the fractional abundance of PBHs calculated for different scalar spectra and latest observational bounds for a monochromatic mass spectrum are displayed. Evidently, in our model, PBHs can be all dark matter in the asteroid-mass window.}
\label{sps}
\end{figure}

In our scenario the parameters which are essentially free to take any value are $\phi_{i}$, $x_0 (\equiv \phi_0/v)$, and any two of the four parameters $a,b,c,d$ while the other two are fixed by the inflection point conditions \eqref{cond1}. Among these parameters, only first two {\it i.e.} $\phi_{i}$ and $x_0$ control the location of the peak in the power spectra and, correspondingly, the mass and abundance of formed PBHs. A significant fractional contribution of PBHs to CDM directly corresponds to a larger height of the bump in the power spectra and also the critical value of the density contrast $\delta_c$ that we choose. After we fix $\delta_c$ for all our calculations, the desired mass fraction is obtained by slightly departing from the conditions for an exact inflection point in the potential. 
In order to explore the feasibility of PBHs production from such a feature in the potential, we resort to a full numerical scan of the parameter space. One should also keep in mind that although the inflaton velocity is exponentially suppressed around the inflection point which is actually required to generate a large peak in the power spectrum, it should still have enough inertia to cross over that region and not get trapped there forever. 

\begin{figure}[t]
\begin{center}
\includegraphics[width=8cm, height=7cm]{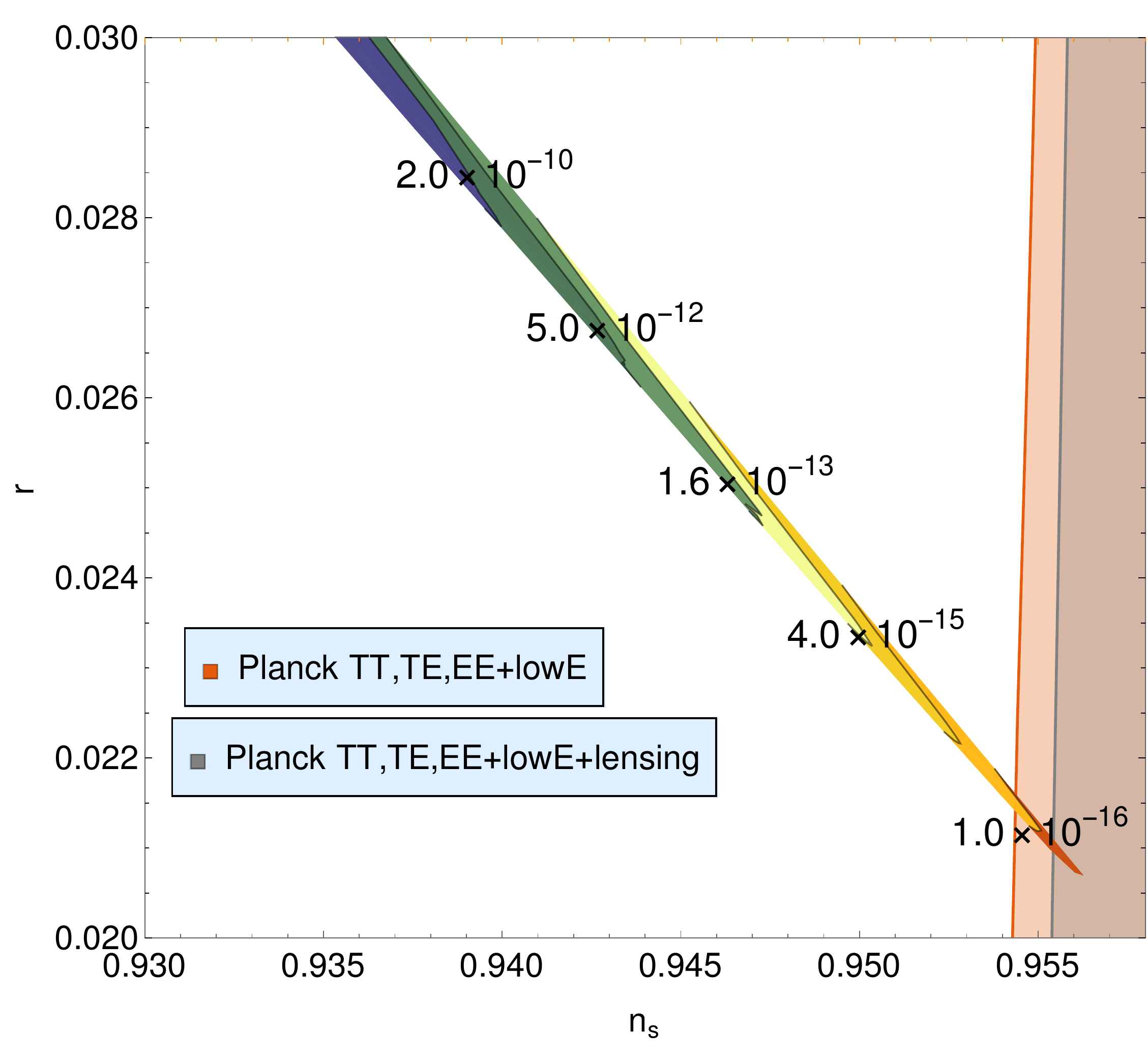}
\caption{The contours of PBHs mass $M_{\rm PBH}/M_{\odot}$ for corresponding allowed values of $n_s$ and $r$ and their comparison with best fit $2\sigma$ contours (shaded regions) from Planck at pivot scale $k_p = 0.002\, {\rm Mpc^{-1}}$ are displayed.}
\label{ns-r-planck}
\end{center}
\end{figure}

We have computed the scalar power spectra numerically for our model with different choices of the model parameters leading to different values of the inflection points as shown on the left in Figure \ref{sps}. In particular, we notice that as we continuously increase the value of $\phi_0$ (the location of the inflection point), the bump in the scalar spectra shifts to the left.  Although the spectra at CMB scales appear nearly flat, we find that there exists a strong correlation between $n_s$ computed at the pivot scale $k_p$ and the position and height of the peak in the power spectrum, and correspondingly, in the mass range and abundance of PBHs. While the value of $n_s$ is consistent with the best fit value from Planck for the spectra on the very right in Figure \ref{sps}, the tension grows as one goes to the left. 
All the spectra show a similar exponential growth corresponding to the scales which leave the horizon around the ultra slow roll phase. We have also computed the exact tensor power spectra for our model. While the scalar spectra show a tremendous rise around the scales of PBHs formation for all the different cases, the tensor spectra show a suppression on the smaller scales but is consistent with a nearly flat spectrum on large scales corresponding to CMB observations. 
We also find a similar trend if we rather change the value of $\phi_i$ {\it i.e.} the scalar spectra shift to the left as we decrease the value of $\phi_i$. 
In Figure \ref{sps} on the right, we have shown the resulting mass fraction obtained from different scalar spectra. Clearly, the power spectrum with a bump on smaller scales leads to a mass fraction in the smaller mass window and vice versa. We also display various exclusion bounds arising from observations in different mass regimes which indicate that PBHs can contribute to the total CDM in two mass windows around $10^{-15} M_\odot$ and $10^{-12} M_\odot$. We also calculate the corresponding values of $n_s$ and $r$ for various spectra and find that lighter mass PBHs are more consistent with the observational constrains than the heavier PBHs. We have shown these contours for $n_s$ and $r$ along with the Planck constraints in Figure \ref{ns-r-planck}.
Following our numerical results, we should further point out that, the same mass fraction can be obtained with different initial conditions either by changing $\phi_{i}$ or by changing $\phi_0$. But, the change in $\phi_{i}$ causes significant variation in the value of $n_s$ at the CMB scales as compared to the case when $x_0$ is changed. In other words, we can use the running of $n_s$ to disentangle the PBH mass fraction arising from these two possibilities.

Before we conclude this section, let us briefly comment on the understanding of the exponential rise in the scalar power spectra arising from the ultra slow roll regime. Recently, it has been discussed in \cite{Byrnes:2018txb} that one can arrive at relevant analytic bounds on the shape of the primordial power spectrum arising in the ultra slow roll regime in the context of single field inflation. In particular, they show that the steepest possible growth has a spectral index of $n_s-1=4$. A detailed calculation by sandwiching an ultra slow roll phase between two slow roll regimes and matching the mode solutions for the curvature perturbations at each boundary, the final spectrum turns out to be $P_{\cal R} \sim k^3 ({\rm log}\, k)^2$ \cite{Byrnes:2018txb} . Interestingly, we find that the growing behaviour of all the spectra that we have plotted in Figure \ref{sps} are perfectly consistent with this logarithmic growth, and not so with the steepest power law growth of the spectrum at the appropriate scales. Note that, this logarithmic correction to the power spectrum comes from the otherwise decaying mode of ${\cal R}_k$ which in this particular situation, becomes a growing mode. Therefore, the weaker logarithmic growth of the power spectrum turns out to be more realistic matching with exact numerical spectra than the steepest growth index for single field inflation. 
Although various causality and analyticity arguments have been used in different contexts to arrive at particular bounds on the growth index of cosmological perturbations \cite{Abbott:1985di}, it seems that these arguments can not be directly applied to the growth of the scalar spectrum associated with the PBH formation as the transient growth only happens for a very short range of modes.


\subsection{PBHs abundance and the effects of reheating}

We shall now turn our attention to understand the effects of an epoch of non-instantaneous reheating on the PBH mass function. 
An epoch of reheating can affect the PBH mass fraction in two different ways:  first, an epoch of slow reheating or an intermediate matter dominated phase with $w=0$ can subsequently enhance the PBH mass fraction \cite{Carr:2018nkm} and second, the presence of a non-instantaneous reheating stage changes the mapping of different length scales from exit to re-entry thereby affecting the normalisation of the scalar power spectra at the pivot scale and hence the mass fraction, even for the monochromatic case, contrary to what has been recently discussed in  \cite{Cai:2018rqf}. In our model inflation lasts upto the horizon exit of reasonably small length scales which minimizes the first effect so we shall mainly focus on the second effect. In order to determine this effect quantitatively, eq. \eqref{a-eq} needs to be modified to include a reheating epoch as 
  \begin{equation}
   \frac{a_{i} }{a_{0}} =\frac{a_{i} }{a_{e} }\frac{a_{e} }{a_{reh}}\frac{a_{reh}}{a_{0}} \qquad {\rm and} \qquad
   \frac{a_{e} }{a_{reh}}=e^{-N_{reh}},
  \end{equation}
where $N_{reh}$ denotes the number of e-folds during reheating. Now,  the temperature $T_r$ at the beginning of the radiation domination depends both on inflation and reheating. Assuming that the reheating epoch is characterised by an equation of state  $w_{reh} =p_{reh}/\rho_{reh}$, we find 
$\rho_{reh} =\rho_{e}\, e^{-3N_{reh}(1+w_{reh})}$.
Upon using $\rho_{reh} = \rho_{r}$ along with \eqref{eq_entrop}, we find the temperature $T_{r}$ at the onset of the radiation epoch as \cite{Cai:2018rqf}
\begin{align}
T_{r} =\left(\frac{90}{\pi^2 g_{*,r}}\right)^{1/4} \left(H_{e}^2 M_{\rm Pl}^2\, e^{-3N_{reh}(1+w_{reh})}\right)^{1/4}.
\end{align}
\begin{figure}[t]
\begin{center}
\includegraphics[width=7.5cm,height=5.4cm]{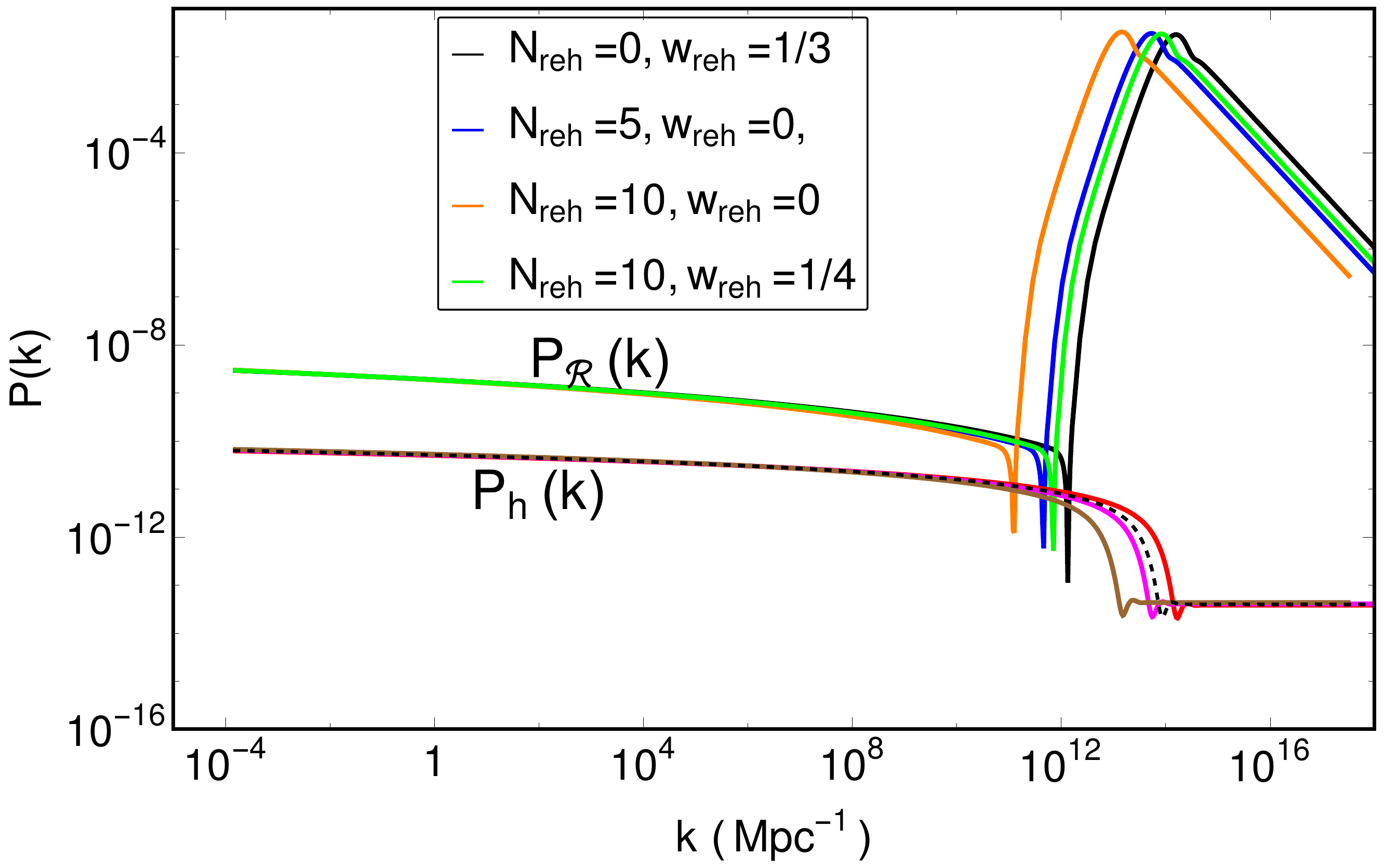}
\includegraphics[width=7.8cm,height=5.4cm]{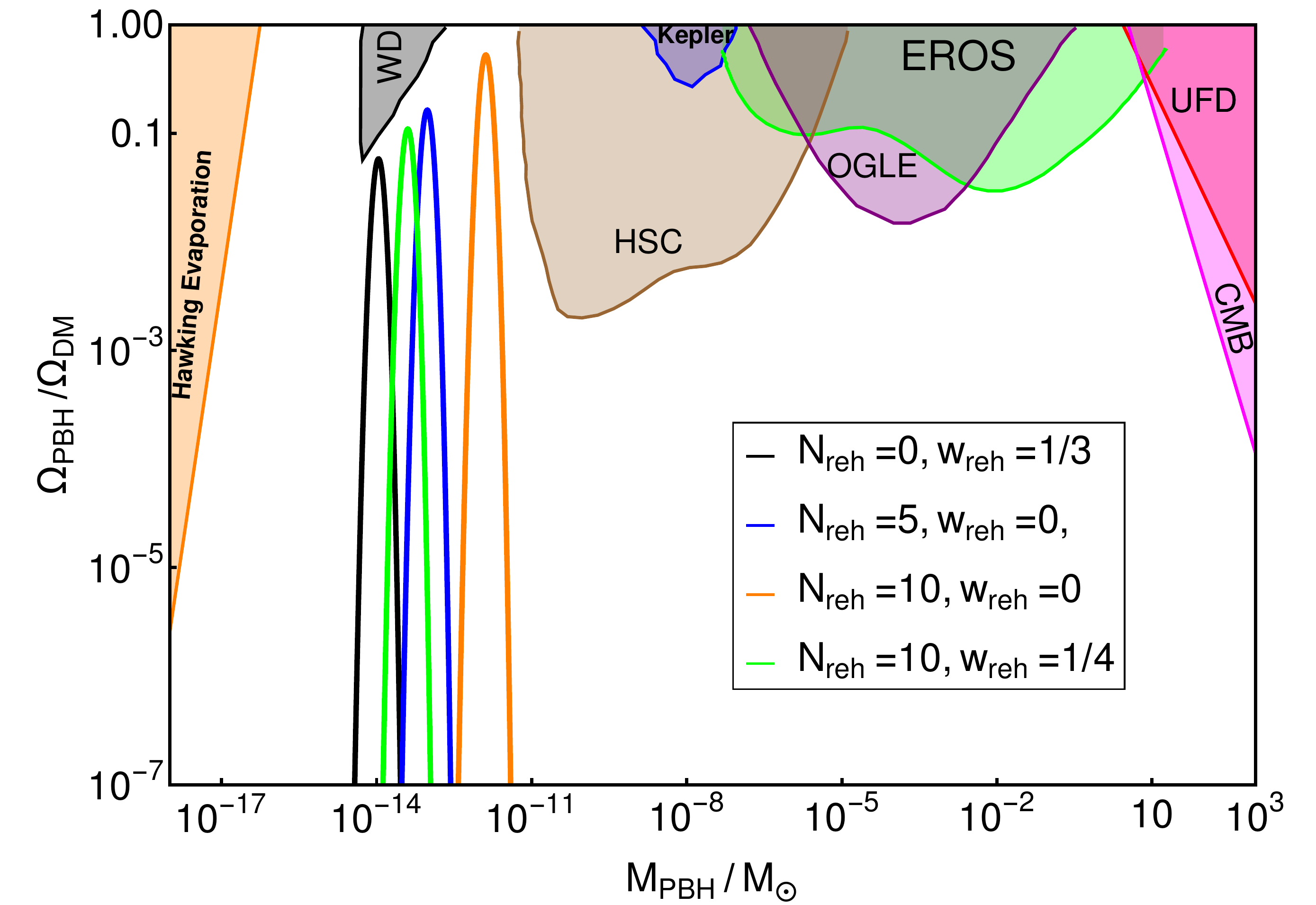}
\caption{On the left, the scalar power spectra $P_{\cal R}$ (solid lines) are shown for our model corresponding to different reheating histories as displayed in the inset. The scalar spectra show a shift to the left for the four different reheating scenarios that we have considered. The tensor spectra $P_h$ for these four cases are also shown with dotted lines. On the right, the PBH mass fraction calculated for these four cases are shown with an increase in the fractional contribution as well as a shift to a larger mass range as the reheating details are changed. Thus a larger mass fraction can be obtained in our model only by changing the reheating details without changing other parameters of the model.}
\label{sps-pbhmf-reh}
\end{center}
\end{figure}
After getting $T_{r}$, we can now use eq. \eqref{a-i0} to obtain the value of $a_i$ to match the observational constraints at the pivot scale  for a given reheating history and then use the same formalism described in the previous section to obtain the power spectra for different reheating histories. It is worth mentioning that the PBH mass fraction exclusively depends on the inflationary power spectrum, if we exclude the first effect of PBH formation during a slow reheating phase, the reheating history plays no role in the determination of the PBH mass fraction from the power spectrum. In our approach, the primordial spectrum contains all the information of the reheating history and the scale $k_{eq}$ remains fixed from observations.  

\begin{figure}[t]
\centering
\includegraphics[width=8cm,height=7cm]{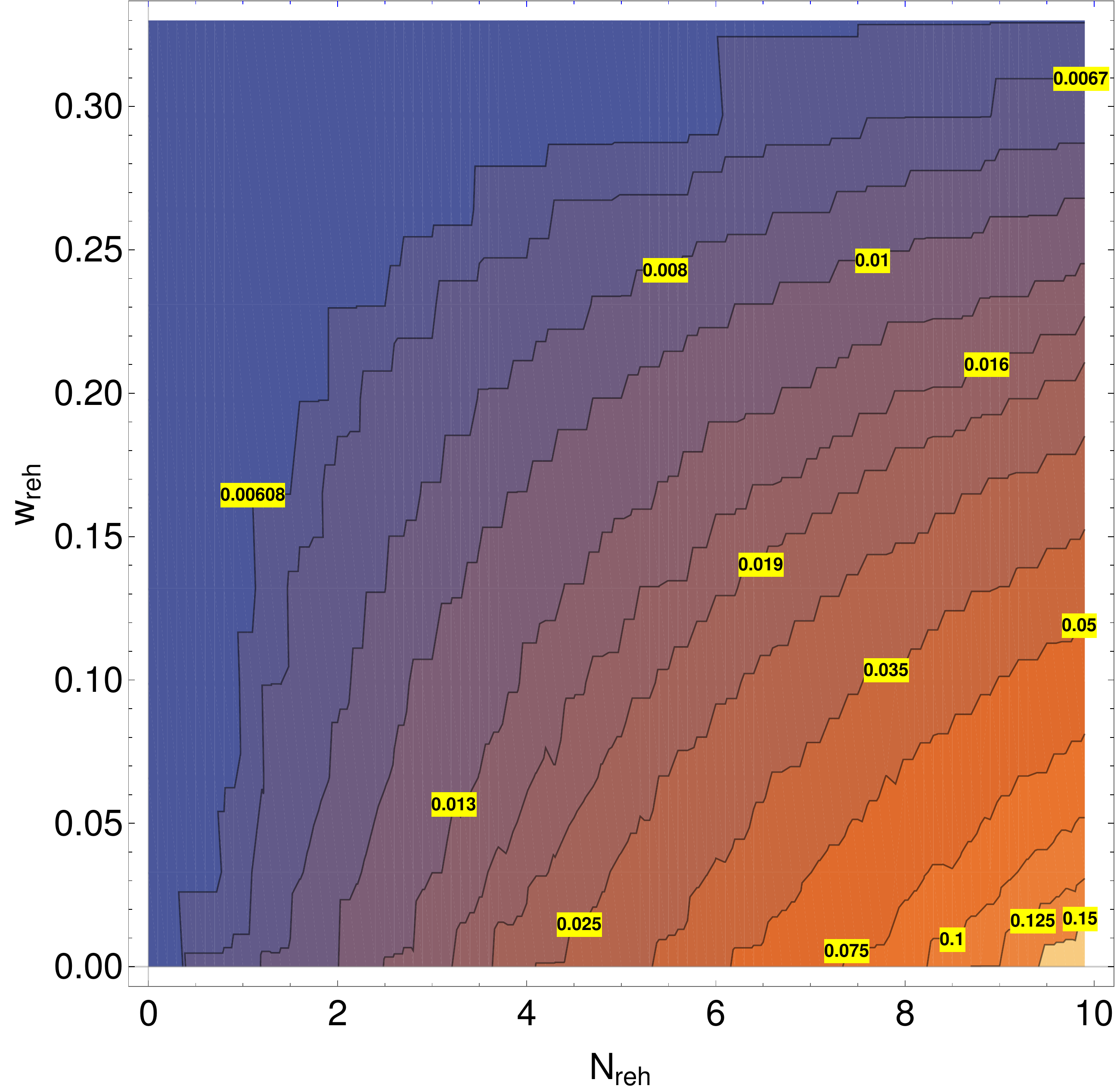}
\vskip 10pt
\caption{The contour plot of $\Omega_{\rm PBH}/\Omega_{\rm DM}$ at equality for different values of $N_{reh}$ and $w_{reh}$ for a fixed value of $\delta_c = 0.414$. The different contours are labelled with their respective fractions $f$ in yellow.  We find that the abundance increases for larger $N_{reh}$ and fixed $w_{reh}$ but it decreases for larger $w_{reh}$ and fixed $N_{reh}$. Our results broadly agree with \cite{Cai:2018rqf} which are derived using the slow roll spectra for a different model.}
\label{reh-contour}
\end{figure}

As a proof of concept, we shall discuss four cases of different reheating histories to show how it affects the scalar power spectra and the PBH mass fraction. In Figure \ref{sps-pbhmf-reh} we have displayed the power spectra and the PBH mass fraction for these four cases wherein the first one with $N_{reh}=0$ and $w_{reh}=1/3$ corresponds to an instantaneous reheating. We find that as the power spectra shift to the left for these different cases, the corresponding PBH mass fraction shift to the right with an increase in the height of the bump thereby leading to a larger contribution of PBHs to the CDM. Since the reheating stage can not be very long in duration, we have restricted to the parameter range $0 \le N_{reh} \le 10$ and $0 \le w_{reh} \le 1/3$.  A substantial increase in the mass fraction happens when  $N_{reh}=10$ and $w_{reh}=0$ {\it i.e.} a matter dominated reheating epoch maximises this effect. In order to gain further insights into the gain in the mass fraction for different reheating histories, in Figure \ref{reh-contour}, we have plotted the PBH abundance $\Omega_{\rm PBH}/\Omega_{\rm DM}$ at equality for different values of $N_{reh}$ and $w_{reh}$ using a contour plot. In particular, we notice that the abundance gets increased for larger value of $N_{reh}$ and fixed $w_{reh}$ while it gets suppressed for larger $w_{reh}$ and fixed $N_{reh}$. These results are in broad agreement with \cite{Cai:2018rqf} which are derived using the slow roll spectra for a different model.  However, it was discussed in \cite{Cai:2018rqf}  that reheating only affects an extended mass fraction of PBHs while a monochromatic mass fraction remains unchanged. Since our model predicts a nearly monochromatic mass fraction for different parameter values, we find that reheating also affects our mass fraction and thus our conclusion is that, an epoch of reheating in general affects the PBH abundance although the effects may not be very significant in some cases. 

Another commonly used reheating parameter $R_{reh}$ is defined as \cite{Ferreira:2013sqa}
\begin{equation}
{\rm log}\, R_{reh}=\frac{-1+3 w_{reh}}{4} N_{reh}.
\end{equation}
Clearly, $R_{reh}=1$ for instantaneous reheating, corresponding to either $N_{reh}=0$ or $w_{reh}=1/3$. Although it is a useful parameter, we find that in our case, it is more convenient to use $N_{reh}$ and $w_{reh}$ and a contour plot as in Figure \ref{reh-contour} gives a much clear picture about the change in the mass fraction with varying these two parameters. Note that $R_{reh} \neq 1$ (with $0 \le w_{reh} < 1/3$) always leads to a larger mass fraction than $R_{reh}=1$ case. 

Before concluding this section, a few remarks about a possible degeneracy arising in our scenario are in order. As shown in Figure \ref{sps}, the scalar power spectra shifts to the left as one increases the value of $\phi_0$ (or the PBH mass fraction shifts to the right as in Figure \ref{sps}). Also, note that, exactly the same shift to the right in the PBH mass fraction happens when the reheating history is changed by varying both $w_{reh}$ and $N_{reh}$ as is evident from Figure \ref{sps-pbhmf-reh}. This indicates that there exists a degeneracy between these two cases as the power spectra and as a result,  the mass fraction could be same for two different choices of parameters. Although tensor power spectra can not break this degeneracy but in principle, very precise measurements of $n_s$ and running of $n_s$ on large scales can possibly break this degeneracy.


\section{Conclusions and discussions}
\label{conclusions}

Inflation provides the right underlying platform to produce the PBHs in the very early universe. While cosmological scales are strongly constrained by the CMB and LSS observations, the small scales existing the horizon during inflation (and re-entering during later epochs) however remain unconstrained. PBHs formation and CMB spectral distortions provide two novel avenues to constrain the small scale dynamics during inflation.  
In this paper, we have studied a single field model of inflation allowing a plateau region for generating PBHs whose potential is broadly motivated from an effective field theory with a suitable flattening at large field values. We should mention here that it is indeed a challenge to find a single field inflationary model with a suitable potential which leads to the desired background dynamics required for producing sufficient abundance of PBHs. While the primordial spectrum on cosmological scales should be consistent with the CMB, the small scale power spectrum should be sufficiently enhanced to produce PBHs. By solving the background and perturbation equations exactly using an adequate numerical scheme, we find that our model can produce a significant abundance of PBHs in different mass ranges. Moreover, our model also generates the entire CDM in PBHs in the asteroid-mass window and the mass fraction turns out to be nearly monochromatic. Note that, in principle, one can obtain a larger mass fraction by choosing a smaller value of $\delta_c$, without changing any other parameters of a given model. However, we also noticed that as we change the values of various parameters in our model, we do generate the PBHs mass fraction in the higher mass range but at the cost of the primordial spectrum being strongly tilted at the CMB scales. 

We have further studied the effects of a reheating phase after the end of inflation onto the mass fraction of PBHs. The reheating epoch is characterised by two parameters $w_{reh}$ and $N_{reh}$ and instantaneous reheating corresponds to $w_{reh}=1/3$ and $N_{reh}=0$ wherein the universe immediately enters into a radiation dominated phase after the end of inflation. We have computed the primordial scalar spectra and the mass fractions for different choices of these two parameters and found that a prolonged matter dominated reheating epoch can lead to very interesting effects on the mass fraction. In particular, we observed that it can shift the mass fraction to larger mass ranges as well as increase the fractional contribution of PBHs to the total CDM. We have computed the effects on the mass fraction for different reheating histories and presented the final results with the help of a contour plot to clearly display the values of $w_{reh}$ and $N_{reh}$ which do lead to the same amount of CDM fraction. We have also calculated the tensor power spectra in our model for different reheating histories. Since the underlying mechanism to produce PBHs during inflation is very model specific, model independent conclusions can not in general be obtained. However, one can still broadly predict the results from two different models if they allow the same background dynamics. 

Finally, we discuss below some future interesting directions to explore further. 
\begin{enumerate}
\item
PBHs can generate a stochastic background of GWs at the smaller scales. Although there exists an upper bound on the tensor-to-scalar ratio $r$ from Planck on the CMB scales, no such bound exists on the scales of the PBHs formation. Since the required amplitude of the power spectrum $P_{\zeta} \sim 10^{-2}$ implying ${\zeta} \sim 0.1$ for abundant generation of PBHs, this implies that the linear treatment of primordial perturbations at those scales is not entirely correct and one should perform a complete non-linear perturbation theory to understand the evolution of these perturbations at such scales. Moreover, the anisotropic stresses induced by the second order perturbations will act as a source term for linear perturbations thereby generating a secondary contribution to the stochastic background of gravitational waves from the epoch of PBHs formation. Of course, such a background will depend on the specific model of PBHs formation. But one can expect that a narrow/broad peak in the scalar spectrum would roughly generate a narrow/broad peak in the GWs spectrum which could then be compared with the sensitivities plots of future GW observatories such as LISA \cite{LISA}.  This has recently been discussed for specific models in \cite{Garcia-Bellido:2016dkw, Garcia-Bellido:2017aan, Guo:2017njn, Sasaki:2018dmp, Clesse:2018ogk}. Induced GWs from PBHs in the mass range around $10^{-12} M_\odot$ correspond to frequency peaked in the mHz range, precisely around the maximum sensitivity of the LISA mission \cite{Bartolo:2018evs}. In addition, there might also exist a GW background from the merger of PBHs in a binary system which could have been created by close encounters of PBHs \cite{Wang:2016ana, Wang:2019kaf}. This possibility and its observational prospects have also been discussed in \cite{Raidal:2017mfl}. It will be interesting to study the constraints arising from these two GWs contributions on our scenario of PBHs formation \cite{future-gw-pbh}. 
\item
In models of PBHs generation, the mass fraction is typically calculated assuming the fact that the initial distribution of primordial curvature perturbations is Gaussian. However, the CMB data from Planck indicates that there exists small amounts of primordial non-Gaussianities (NG) \cite{Ade:2015ava}. It has already been discussed how such primordial NGs would affect the mass fraction of PBHs and in some cases, the resulting effects could be very significant \cite{Hidalgo:2007vk, Young:2013oia, Young:2015cyn, Franciolini:2018vbk, Atal:2018neu}. Further, there could also arise additional NGs due to the generation of a bump in the power spectrum which could affect the PBHs formation at those scales. Therefore, in all these models, one can use the abundance of PBHs as a tool to constrain NGs on very small scales which can not otherwise be constrained \cite{Byrnes:2012yx}.   In addition, the non-linear relation between the curvature and density perturbations introduces significant NGs in the over-density statistics even when the curvature perturbation has an exactly Gaussian distribution. It has been recently discussed that the abundance of PBHs is very sensitive to such non-linear effects \cite{DeLuca:2019qsy, Young:2019yug, Kehagias:2019eil}. Moreover, there could also arise a dominant contribution to the background of induced GWs from non-Gaussian primordial scalar perturbations \cite{Cai:2018dig, Unal:2018yaa}. Since these imprints could be very model dependent, it will be worth exploring them in our model arising both from the bispectra and trispectra as well as from the intrinsic non-linearities. 
\item
The calculation of the PBHs mass fraction in a given model is generally done in a classical set-up. However, it has been greatly discussed that quantum diffusion during the ultra slow roll phase around the inflection point would be very relevant since the inflaton field in this region undergoes a phase of strong deceleration. In single field slow roll inflation, the classical evolution of the field fluctuations is $\delta \phi_{\rm Cl}\sim \dot \phi/H$ while the quantum evolution is given by $\langle\delta \phi_{\rm Q}\rangle\sim H/2\pi$. The classical evolution leads to correct predictions for the power spectrum when $\delta \phi_{\rm Cl} >  \langle\delta \phi_{\rm Q}\rangle$. However, there could be a situation when this condition does not hold true. In particular, this happens when $\dot \phi$ becomes extremely small {\it i.e.} during the ultra slow roll phase which is very crucial for PBHs generation. It has been pointed out that the classical calculation of PBHs mass fraction during this phase does not necessary give the correct results and one should take into account the contributions arising from the quantum diffusion \cite{Biagetti:2018pjj, Ezquiaga:2018gbw, Kuhnel:2019xes}. The resulting effects on the PBHs mass fraction due to quantum diffusion will depend on the underlying background dynamics of an inflationary model and it will be very important to calculate this in our model. We leave the computation of quantum diffusion in our model and its implications for future work \cite{future-qf-pbh}. 
\item
Super massive black holes (SMBH) are observed at the centre of galaxies and in quasars at high redshifts, $z \sim 6-7$ \cite{Mortlock:2011va, Reed:2019ftq}. On the contrary, it is very challenging to form such SMBHs in the standard $\Lambda$CDM cosmology. It has been pointed out that the tail of the PBHs distribution can serve as the seeds for the origin of such SMBHs. PBHs produced with masses larger than $10^4 M_{\odot}$ at around $z\sim 15$ can provide seeds which can then accrete matter over time and probably merge by today to form these SMBHs \cite{Clesse:2015wea}. It has recently been shown that Eddington accretion would be enough for explaining SMBHs today and super-Eddington accretion is not necessary required \cite{Pandey:2018jun}. Furthermore, intermediate PBHs can also be produced in the same model which can explain the presence of ultra-luminous X-ray sources \cite{Liu:2013jwd, Bachetti:2014qsa}. Since our scenario predicts PBHs in all mass ranges, it will be interesting to analyze whether the tail distribution of PBHs in our model can give rise to seeds for the SMBHs. 

\end{enumerate}


\section*{Acknowledgment}
We wish to thank Kanhaiya Lal Pandey and Ranjan Laha for useful discussions. We also thank an anonymous referee whose comments and suggestions have improved the overall presentation of the paper. RKJ would like to acknowledge the financial support from the new faculty seed start-up grant of IISc and partial support from the Science and Engineering Research Board, Department of Science and Technology, Government of India, through the Core Research Grant CRG/2018/002200.

\appendix
\section{Background evolution: slow roll, ultra slow roll and all that}
\label{app-A}

After specifying the potential, we can now study the background inflationary dynamics resulting from this potential.
Using the number of e-folds, $N(t) = {\rm ln}\, \l(a(t)/a_i\r)$, as the independent time variable, the system is governed by the following Friedmann equations 
\be
H^2 &=& \frac{V(\phi)}{M_{\rm Pl}^2 (3-\epsilon)},\\
\frac{d H}{d N} &=&  - \frac{H}{2 M_{\rm Pl}^2} \left(\frac{d \phi }{d N}\right)^2,\label{frd-2}
\ee
with the Klein-Gordon equation for $\phi$ as
\be
\frac{d^2 \phi}{d N^2} + (3-\epsilon) \frac{d \phi}{d N} + \frac{1}{H^2}V'(\phi) =0, \label{frd-1}
\ee
where $\epsilon$ is the first Hubble slow roll parameter, given by
\beq
\epsilon =   -\frac{\dot H}{H^2} = \frac{1}{2 M_{\rm Pl}^2} \left(\frac{d \phi}{dN } \right)^2. \label{fsp-1}
\eeq
We also define the second slow roll parameter $\eta$ as 
\beq
 \eta = -\frac{\ddot \phi}{H \dot \phi} = \epsilon - \l(\frac{d^2 \phi/{d N^2}}{d \phi/d N}\r)
 \eeq
Of course, only two out of the above three equations are independent. To solve these equations completely, we need three initial conditions: $\phi_{i}$, $d\phi_i/d{N}$ and $H_i$. Since the evolution of the field $\phi$ starts in the slow roll regime at large field part of the potential, after choosing an appropriate value of $\phi_{i}$, the values  of $d\phi_i/d{N}$  and $H_i$ can be chosen from the slow roll conditions and the Friedmann equations. In slow roll approximation wherein $\frac{d^2 \phi}{d N^2} \ll  \frac{d \phi}{d N}$ and $\epsilon \ll 1$, eq. \eqref{frd-1} reads
\beq
\frac{d \phi}{d N} + \frac{1}{3 H^2}V'(\phi)  \simeq 0, 
\eeq
whose solution is given by
\beq
\phi(N) \simeq \phi_i -\sqrt{2\epsilon_{_V}} M_{\rm Pl} (N-N_i),
\eeq
where $\epsilon_{_V}$ is the first potential slow roll parameter defined as $\epsilon_{_V} \equiv \frac{M_{\rm Pl}^2}{2}\l(\frac{V'}{V}\r)^2$. It is clear form this slow roll solution that $\phi$ is a monotonically decreasing function of $N$. However, the same does not remain valid in the regime of ultra slow roll wherein the slow roll conditions do not hold. Since the potential is very flat and $\epsilon$ becomes very small, one can now ignore the last term in \eqref{frd-1} and thus it reduces to 
\beq
\frac{d^2 \phi}{d N^2} + 3 \frac{d \phi}{d N} \simeq 0, 
\eeq
which leads to 
\beq
\frac{d \phi}{d N} \sim {\rm exp}\, [-3(N-N_i)], 
\eeq
and therefore, the inflaton velocity gets exponentially suppressed. This also implies that $\epsilon \sim {\rm exp}\, [-6(N-N_i)]$ and $\eta \simeq \epsilon + (3- \epsilon) \sim 3$ during the ultra slow roll phase which is also evident in Figure \ref{3_back}. As we have discussed in Sec. \ref{bgnd}, this behaviour leads to an exponential enhancement of the curvature perturbations which induces a tremendous growth of the power spectrum around the scales corresponding to the ultra slow roll regime.

\section{PBHs mass fraction and associated uncertainties}
\label{app-B}

It is well known that the calculation of the PBHs mass fraction for a primordial power spectrum $P_{\cal R} (k)$ suffers from many approximations and uncertainties and each of these uncertainties changes the final result drastically.  As we shall discuss here, the PBH mass fraction needs to be in a very narrow mass range to lead to a significant contribution to the total CDM and therefore, in order to produce such a monochromatic mass spectrum, the choice of the collapse formalism or the value of the critical density contrast becomes very crucial. In what follows, we shall briefly discuss some of these uncertainties. 

\subsection{Peaks theory vs. Press-Schechter formalism}

\begin{figure}[t]
\begin{center}
\includegraphics[width=5.8cm,height=6.8cm]{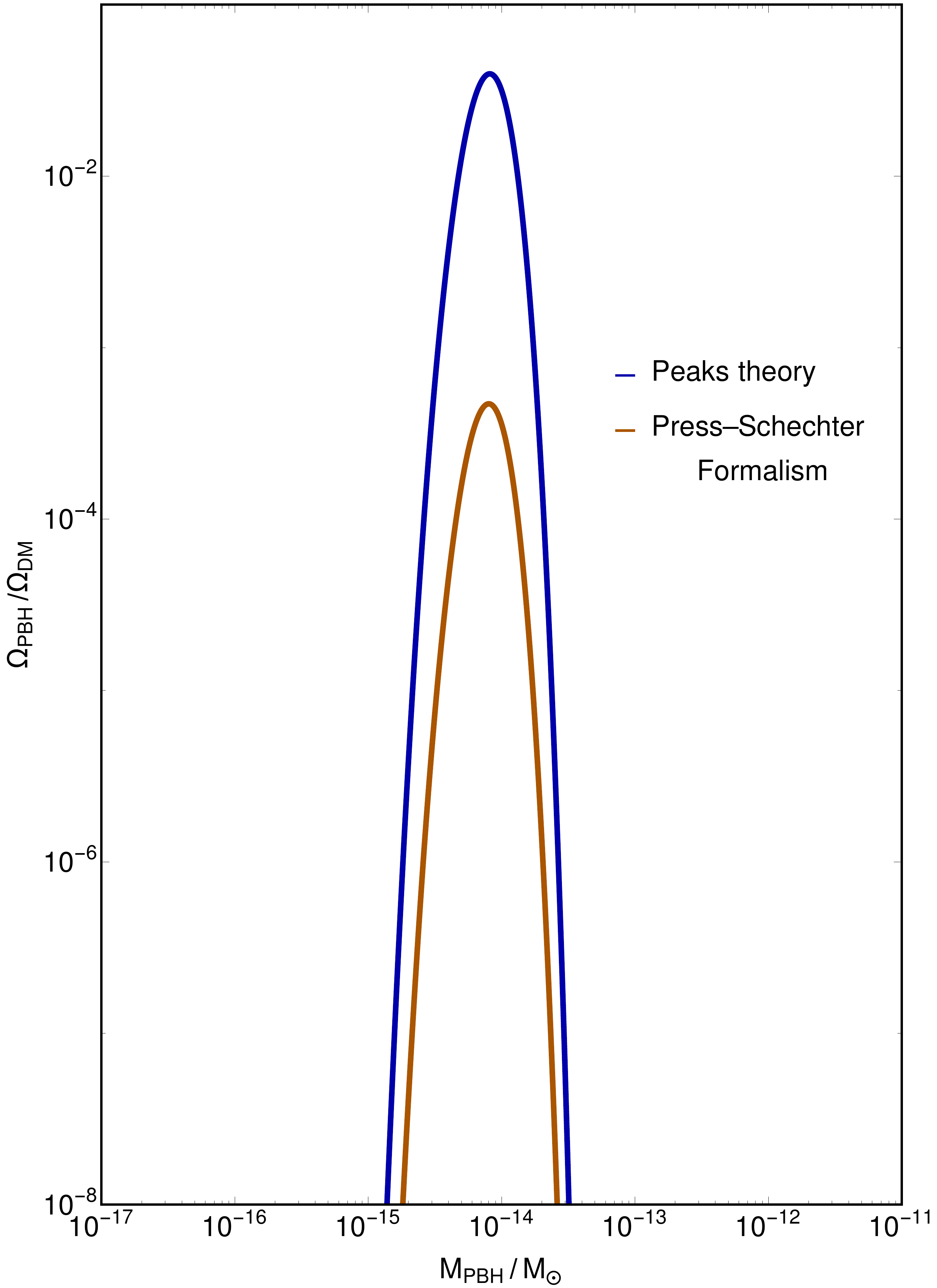}
\hspace {20pt}
\includegraphics[width=5.8cm,height=6.8cm]{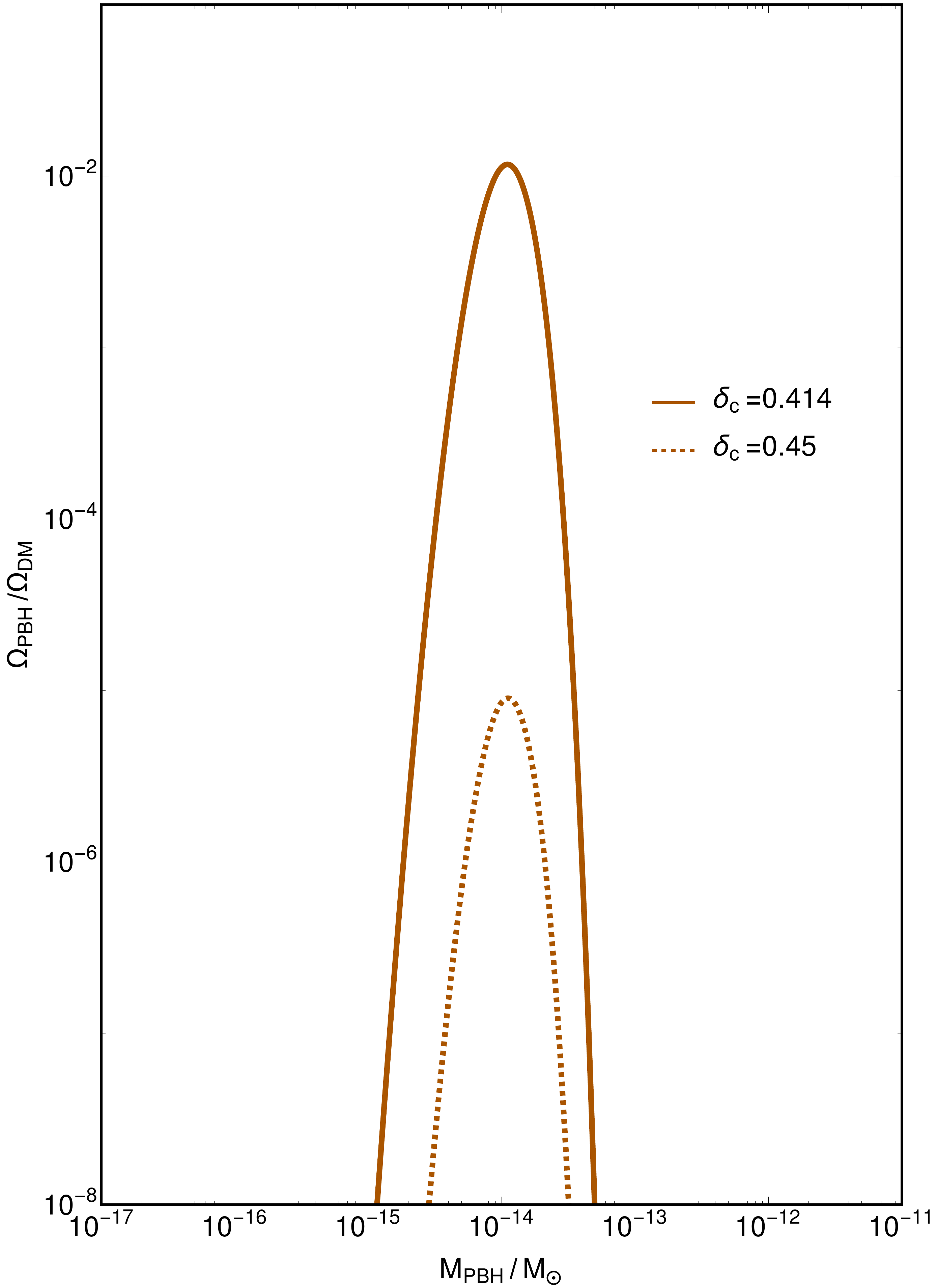}
\vskip 10pt
\caption{On the left, the mass fraction of PBHs at the radiation-matter equality has been plotted as a function of the mass of PBHs for two different formalism: Peaks theory and Press-Schechter formalism. It is evident that the peaks theory predicts a larger mass fraction than the Press-Schecter formalism for the same choice of parameter with the same primordial spectrum $P_{\zeta}(k)$. On the right, the exponential sensitivity of the PBHs mass fraction on the value of the critical density contract $\delta_c$ in the Press-Schechter formalism is displayed. Clearly, a tiny change in $\delta_c$ leads to a significant change in the abundance of PBHs.}
\label{press-peak}
\end{center}
\end{figure}

Primordial overdensities on smaller scales collapsing to form PBHs instantaneously right after the horizon entry of these scales in the radiation dominated universe can be described by two formalisms: Peaks theory and Press-Schechter.  Both the formalisms use a critical overdensity above which the fluctuations should collapse and form PBHs. In Peaks theory, the critical value is stated in terms of the peak value of a fluctuation while in the Press-Schechter approach, it is calculated as the average value of a fluctuation. In principle, the relationship between the peak value and the average value of a fluctuation depends on its shape but in practice, they are expected to differ only by a factor of order unity, with the peak value being higher. In general, results from these two established methods  do not show convergence with all the other criterias and parameters fixed \cite{Young:2014ana,Musco:2018rwt,Yoo:2018kvb}. For our model which produces a nearly monochromatic PBH mass fraction,  we find that the calculated mass fraction from peaks theory is somewhat higher than the Press-Schechter formalism as shown in  the left panel of Figure \ref{press-peak}. However, the PBH mass range converges with good accuracy in both cases. 


\subsection{Critical value of the density contrast} 

A critical threshold value of the density contrast $\delta_c$ plays an extremely crucial role in understanding the collapse and formation of PBHs. 
It has been discussed at length that an appropriate parameter that should be used to calculate the PBH abundance is the density contrast $\delta$ \cite{Young:2019osy}. 
However, often the critical value of a metric perturbation such as the curvature perturbation $\zeta_c$ has also been used extensively to determine the PBHs mass fraction \cite{Shibata:1999zs}. Such a choice simplifies the calculations of the mass fraction but, as pointed out in \cite{Young:2014ana}, $\zeta_c$ is strongly dependent on the local environment unlike $\delta_c$ which was also numerically verified in \cite{Harada:2015yda}. Thus, the density contrast should be considered a more natural (and physical) parameter to describe to collapse criteria.  Since the environmental effects are likely to change the collapse threshold in terms of $\zeta_c$ which will then affect the resulting mass function, we shall use $\delta_c$ in all our calculations to avoid this.

Until now, we lack the knowledge of a very specific value of $\delta_c$ above which the PBHs should form \cite{Shibata:1999zs,Harada:2013epa}. Using the Jeans collapse condition as the criteria for the PBH formation, the first order of magnitude estimate of the density contrast was provided as follows \cite{Carr:1974nx}
\beq
w\simeq\delta_c<\delta<\delta_{max}\sim 1.
\eeq
Here, $w$ is the equation of state parameter described by $w=p/\rho $. Since $w=1/3$ during the radiation dominated phase, 
$\delta_c \simeq 1/3=0.33$. 
The upper limit on $\delta_c$  was initially set by Hawking and Carr from separate universe scenario \cite{Carr:1974nx, Carr:1975qj} but it was realized much later by Kopp, Hofmann and Weller that this limit is in fact a geometrical consequence and not necessarily a result of the separate universe approach \cite{Kopp:2010sh}.
Subsequent works to find out the lower limit on $\delta_c$ were based on approaches using numerical relativity \cite{Shibata:1999zs, Musco:2008hv, Baumgarte:2016xjw}. In Ref. \cite{Shibata:1999zs}, the critical value of $\zeta$ was obtained as $\zeta_c=1.4-1.8$ which corresponds to $\delta_c=0.3-0.5$.
Later, Polnarev and Musco did this analysis for $\delta_c$ and converged upon $\delta_c =0.45-0.66$ \cite{Polnarev:2006aa}.
This numerical result was also supported by an analytical formula obtained a few years later which suggested  $\delta_c = 0.414$ for the radiation dominated era \cite{Harada:2013epa}. Further, this analytical estimate of $\delta_c$ is in rough agreement with the range specified from the numerical analysis in \cite{Shibata:1999zs}. For the scenario that we consider in this paper,  it is evident that the numerically allowed range of $\delta_c$ leaves a huge space for uncertainty in the calculation of mass fraction which is also evident in the right panel of Figure \ref{press-peak}.  Therefore, to avoid ambiguity, we shall use the analytical value suggested for critical density contrast $\delta_c = 0.414$ throughout our analysis. 
Note that, the value of $\delta_c$ also depends on the shape of the inflationary power spectrum but we shall neglect that effect for simplicity \cite{Germani:2018jgr}. Recently, it has also been discussed that the abundance of PBHs is quite sensitive to non-linear effects such as the non-linear relation between $\zeta$ and $\delta$ \cite{DeLuca:2019qsy, Young:2019yug, Kehagias:2019eil}.


\subsection{Choice of the window function}

In order to estimate the PBH abundance in a given inflationary model, one must relate the primordial power spectrum in Fourier space to the probability distribution function in the real space by coarse-graining the curvature perturbations with the help of a window (smoothing) function. It has been noted that different choices of a window function leads to a different relation between the PBH abundance and the required power spectrum and therefore, one typically does not have a one-to-one correspondence between the two as has been extensively discussed in \cite{Ando:2018qdb, Musco:2018rwt, Young:2019osy}. 
In practice, the widely used window functions are (i) the real space top-hat window function, (ii) Fourier space Gaussian window function, and (iii) Fourier space top-hat window function. The choice of a window function thus turns out to be an important factor that can change the predictions of the collapse criteria drastically. For our calculations in this paper, we shall use the most commonly used Fourier space Gaussian window function which is given by
\beq
W(k,R)={\rm exp}\l(-\frac{k^2R^2}{2}\r),
\label{window}
\eeq
where $R$ is the scale associated with the smoothing of primordial overdensities.

\bibliographystyle{JHEP}
\bibliography{bibfl1}
\end{document}